\documentclass[aps,10pt,singlecolumn,superscriptaddress,amsmath,amssymb]{revtex4-2}
\usepackage[utf8]{inputenc}
\usepackage{subfiles}
\usepackage{graphicx}
\usepackage[colorlinks,allcolors=cyan!70!black]{hyperref}
\usepackage{float}
\usepackage[normalem]{ulem}
\usepackage{comment}
\usepackage{comment}
\usepackage{lipsum}
\usepackage[dvipsnames]{xcolor}
\definecolor{deepskyblue}{rgb}{0.0, 0.75, 1.0}

\begin{document}

\preprint{APS/123-QED}

\title{Single-cell-level distributions and relationships can differentiate cell-division and growth models}

\author{Vikas}
 \email{vikaskaushikkarora@gmail.com}
 \affiliation{Department of Physics, Indian Institute of Technology Delhi, Hauz Khas, 110016, New Delhi, India}
\author{Rahul Marathe}
 \email{maratherahul@physics.iitd.ac.in}
 \affiliation{Department of Physics, Indian Institute of Technology Delhi, Hauz Khas, 110016, New Delhi, India}
\author{Anjan Roy}
 \email{anjanroy@dbeb.iitd.ac.in}
 \affiliation{Department of Biochemical Engineering and Biotechnology, Indian Institute of Technology Delhi, Hauz Khas, 110016, New Delhi, India}

\date{\today}

\begin{abstract}
Complex interactions among regulatory molecules determine the rules underlying cell growth and division in microbial cells. While the governing molecular network may not always be obvious, it is well known that correlations among certain physiological quantities measured in experiments, such as birth-size, division-size, division-time, and division-added-size, can differentiate among various cell-division models, such as Timer, Sizer, and Adder. Here we show that, apart from these correlations, which we extend for the case of stochastic single-cell growth and stochastic asymmetric partitioning, probability distributions of these quantities and statistical relationships between them can also be used to differentiate between these division models. Interestingly, we show that these quantities can not only differentiate the division models, but also distinguish among the single-cell growth paradigms, such as linear and exponential growth. We then demonstrate this differentiability among various division and growth models by comparing our analytical results with published experimental data. We further show that these results remain valid even when the growth rate of a cell is correlated with the growth rate of cells from previous generations in the lineage.

\end{abstract}

\maketitle

\section*{Statement of Significance}
Different growth and division strategies in bacterial species imply different underlying regulatory mechanisms, and identifying the correct strategy helps pinpoint the correct molecular network that controls cell size and division. This also aids in identifying evolutionary strategies or constraints in different microbial species. Additionally, the determination of the type of single-cell growth paradigm, such as linear and exponential, is relevant in order to better understand the dynamics of microbial systems. In this study, we obtain analytical results concerning various single-cell quantities, viz. birth-size, division-size, division-time, and division-added-size, and show that they can be used to distinguish between various division models, such as Sizer and Adder, as well as between different growth modes, such as linear and exponential. 

\section{Introduction} \label{sec: introduction}
Understanding bacterial cell growth and division is crucial in microbiology, biotechnology, and biochemical engineering. Despite the underlying complexity of these processes, extensive research conducted over several decades has revealed that bacterial cells often follow simple, yet relatively robust principles, such as the so-called bacterial growth laws and size laws \cite{schacMaloe1958, schacMaloe1958B, scottMatthew2010, scottMatthew2014, donachie1968, helmstCooper1968}. In particular, it was found that cell division is governed by regulatory mechanisms that can be effectively captured using coarse-grained models. Earlier models of bacterial cell division suggested that cells follow either the Timer or the Sizer mechanism \cite{powell1956, kochSchac1962, powell1964, wheals1982, diekmann1983, tysonDiekmann1986}. The Timer model posits that the cells in the culture divide after a fixed time from birth, implying that division is governed purely by age \cite{powell1956}. In contrast, the Sizer model suggests that division occurs once a cell reaches a critical size, ensuring a threshold for division \cite{kochSchac1962, powell1964, wheals1982, diekmann1983, tysonDiekmann1986}. While the Sizer model gained early traction, Koppes et.al. criticized the model based on various inferred correlations and the skewness in division-time, which did not agree with the model's predictions \cite{koppes1980, voornKoppes1997}.

Although some microorganisms may indeed follow the Sizer division strategy \cite{facchettiChang2017, miottoGosti2024, nietoGarcia2020}, recent experimental studies on various bacterial species \cite{santiDhar2013, campos2014, taheriJun2015, fievetDucret2015, deforetDitmarsch2015, priestman2017, chungKarAmir2024, messelink2021, logsdonAmir2017}, including \textit{Escherichia coli}, \textit{Bacillus subtilis}, \textit{Caulobacter crescentus}, \textit{Mycobacterium smegmatis}, \textit{Mycobacterium tuberculosis}, \textit{Desulfovibrio vulgaris}, and \textit{Corynebacterium glutamicum}, have shown that these species instead follow a third strategy, known as the Adder principle, or the Adder model \cite{voornKoppes1997, amir2014, junReview2015, junReview2016}. According to the Adder principle, a cell divides after adding a constant amount of biomass, regardless of its initial size or age. Besides bacterial cells, yeasts and mammalian cells also follow division strategies close to the Adder model \cite{chandlerBrown2017, soiferAmir2016, cadartGrilli2018}.

\begin{figure}[h!]
    \centering
    \begin{tabular} {cc}
        \includegraphics[width=0.47\textwidth]{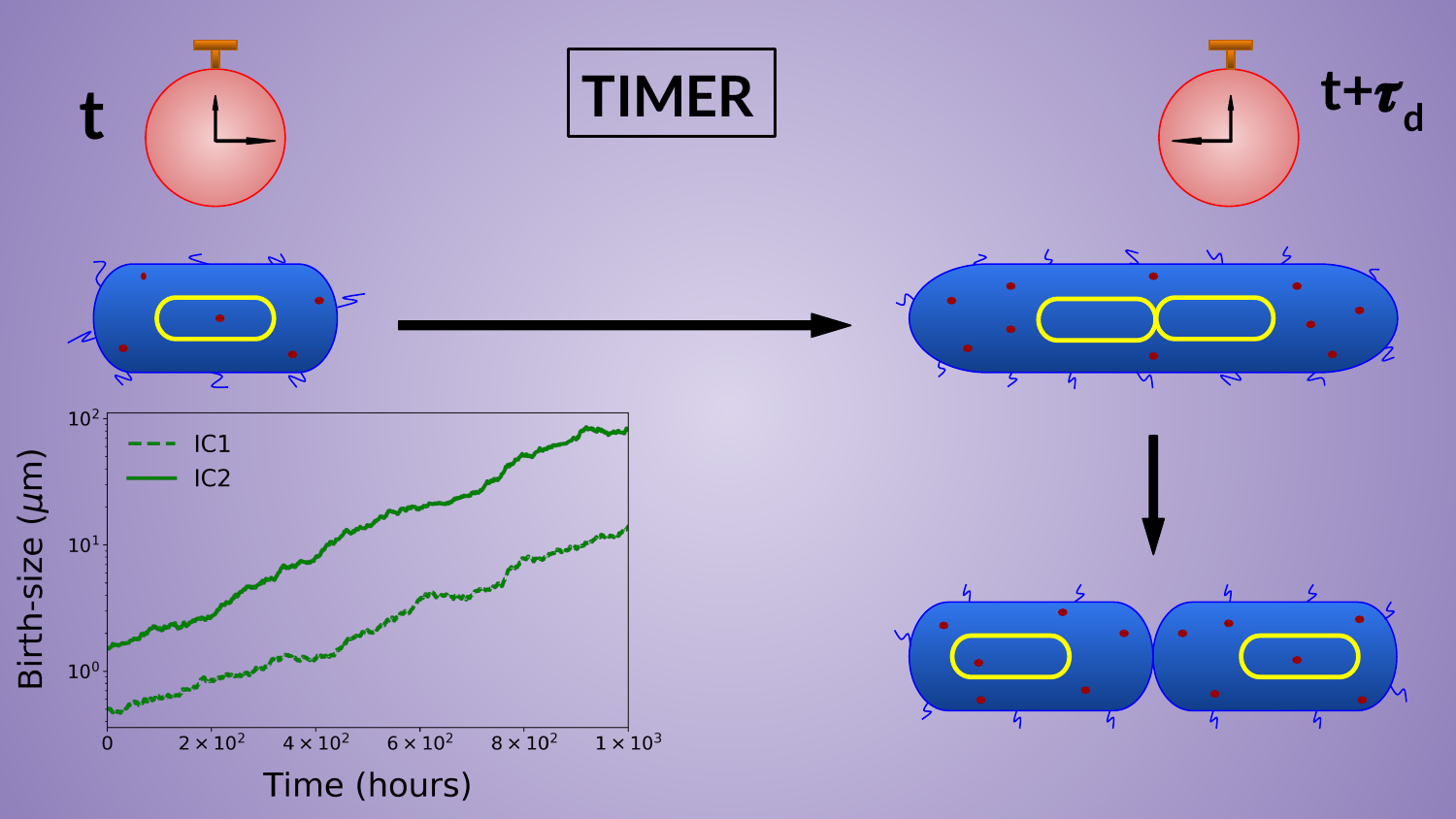}&
        \includegraphics[width=0.47\textwidth]{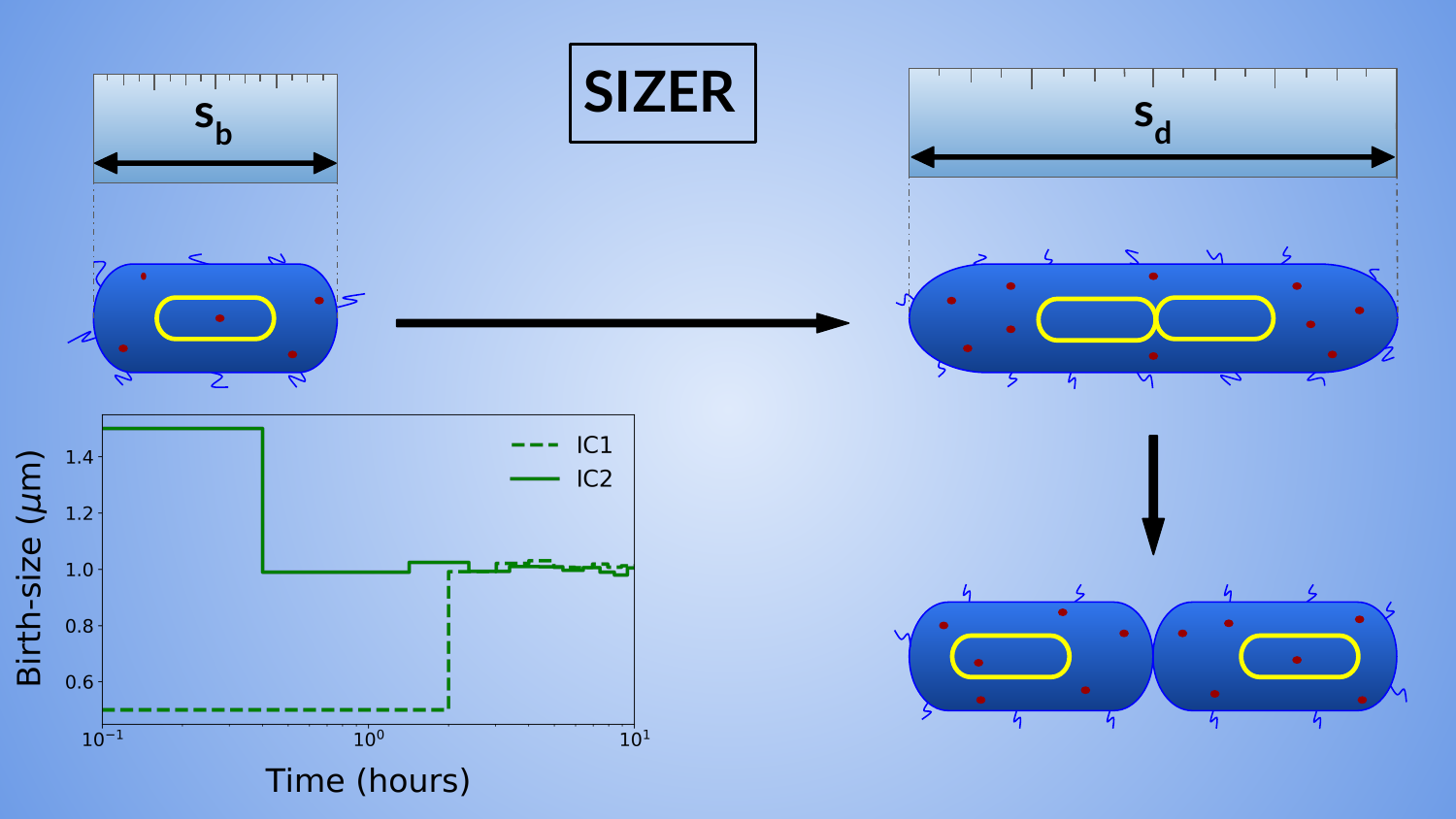}\\
    \end{tabular}
    \includegraphics[width=0.47\textwidth]{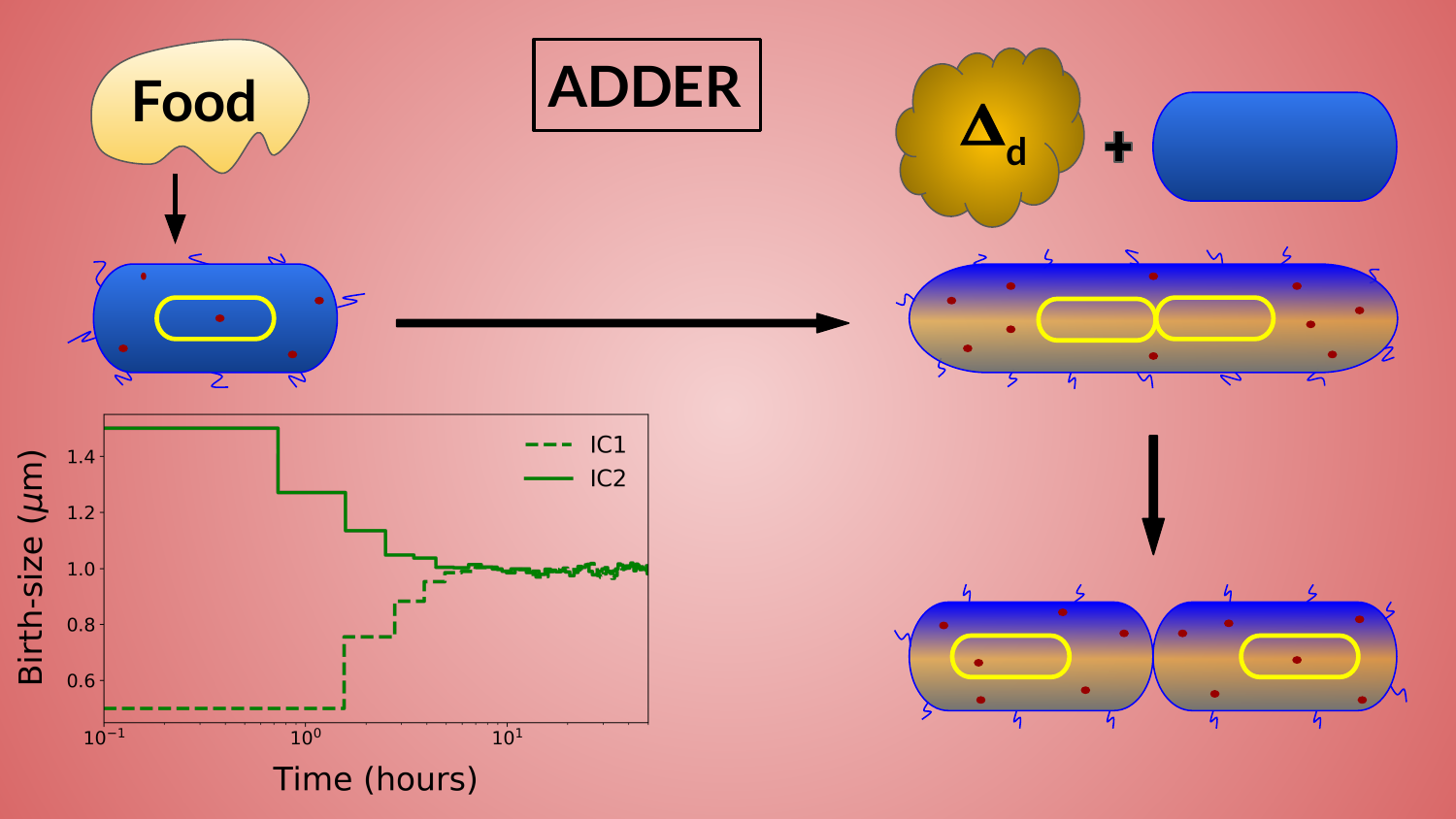}\\
    \caption{Artistic representations of various cell-division models: Timer, Sizer, and Adder. The insets represent the graphs of birth-size of a random cell (versus time) tracked for many generations that follows a particular cell-division model under exponential growth, with stochasticity in the rule of cell-division, growth rate, and partitioning fraction. Here, for simplicity, the distributions for the free parameters are taken to be Gaussian. The mean division-time is taken to be $1$ hour for the Timer model, mean division-size is taken to be $2 \mu$m for the Sizer model, and mean division-added-size is taken to be $1 \mu$m for the Adder model. Mean exponential growth rate is taken to be $\ln(2)$ per hour and mean partitioning fraction is taken to be $0.5$. The coefficient of variation for each of these distributions is taken to be $1$ \%. IC1 and IC2 represent `initial condition 1' (initial birth-size being 0.5) and `initial condition 2' (initial birth-size being 1.5), respectively.}
    \label{fig: cartoon}
\end{figure}

It is interesting to note that while the Sizer and Adder rules enable cell-size homeostasis, Timer does not \cite{junReview2015, junReview2016, willisHuang2017, vuaridelDhar2020}. Cell-size homeostasis is fundamental to bacterial survival, ensuring cells maintain a stable birth-size over generations. A bacterial cell cannot be too small because a minimum cellular volume is required to house the essential molecular machinery of life, maintain viable intracellular concentrations of key metabolites against stochastic fluctuations, and support a mechanically stable cell membrane capable of sustaining vital membrane-associated functions \cite{pirie1973}. Also, bigger cells face other problems, such as larger diffusion time for essential metabolites, a lower surface-to-volume ratio emphasizing imbalances between substrate supply and metabolic demand, and insufficient scaling of biosynthetic capacity with cell volume resulting in disruption of intracellular biochemical organization \cite{marshallYoung2012, dillGhosh2011, amodeoSkotheim2016, neurohrTerry2019}. Therefore, the cells should have an optimal size to operate appropriately, and maintaining cell-size homeostasis ensures that future generations of cells will continue to be in their optimal size.

The Sizer model naturally maintains cell-size homeostasis by enforcing division at a fixed size. Regardless of initial birth-size, subsequent newborn cells are always half the division-size, particularly for symmetrically dividing cells, ensuring a stable birth-size distribution over time. Similarly, the Adder model achieves homeostasis by requiring cells to grow by the addition of a fixed biomass before division. Any size variation at birth diminishes over the successive divisions, leading to convergence toward a stable birth-size. This property of cell-size homeostasis for the Sizer and Adder models can also be inferred from the insets in Fig. \ref{fig: cartoon}, where the average birth-size approaches a constant value with time. However, the Timer model, in general, does not ensure homeostasis. For exponential single-cell growth, homeostasis is maintained only for fine-tuned values of the growth rate, partitioning fraction, and division-time. Stochasticity in these parameters further causes the birth-size in a cell-lineage to diverge over successive generations (See Appendix \ref{subsec: SbirthSizeTimer}). This behavior is evident from the inset corresponding to the Timer model in Fig. \ref{fig: cartoon}, where the birth-size continues to increase with generation number, demonstrating the absence of size homeostasis. Interestingly, for linear single-cell growth, the Timer model effectively becomes the Adder model \cite{willisHuang2017, vuaridelDhar2020} (See Appendix \ref{sec: StimerLinear}), and consequently exhibits cell-size homeostasis.

Different cell-division strategies, including the Timer, Sizer, and Adder models, are associated with distinct biochemical reaction-network mechanisms that regulate the timing and coordination of cell division. Several such mechanisms have been identified, for example, in \textit{Schizosaccharomyces pombe} and \textit{Saccharomyces cerevisiae}, where activator accumulation and inhibitor dilution mechanisms lead to the Sizer-like behavior for the growth phase and Adder-like behavior for the full cell-cycle \cite{marshallYoung2012, facchettiChang2017, chandlerBrown2017}. Additionally, some other size-sensing mechanisms that determine the timing of replication initiation and cell division, have also been discovered for bacterial species, such as \textit{E. coli} and \textit{B. subtilis}  \cite{robert2015, willisHuang2017}. Further, it is proposed that the Adder model is a direct consequence of the fact that cells accumulate specific division-related proteins and precursors up to a fixed threshold number before committing to division \cite{siJun2019, pandeyJain2020, nietoGarcia2024}. Knowing which division model is followed by a specific bacterial species can be useful to pinpoint the underlying mechanism for cell-size control and understand its evolutionary relation to other species. 

Although it is known that the different models of cell divisions can be distinguished by the correlations between single-cell quantities, such as birth-size, division-time, division-size, and division-added-size \cite{nietoGarcia2020, santiDhar2013, campos2014, taheriJun2015, deforetDitmarsch2015, fievetDucret2015, priestman2017, junReview2015, junReview2016, soiferAmir2016, chungKarAmir2024, messelink2021, logsdonAmir2017, chandlerBrown2017, cadartGrilli2018}, in this study, we extend these results for the case of stochastic growth rate (over individual cells in a lineage), asymmetric division, and partitioning stochasticity. Additionally, we obtain the analytical form of the distributions of these above-mentioned quantities and show that they can be used to differentiate between the division models. We further obtain statistical relationships between these single-cell quantities and show that some of these relationships can also differentiate between the Sizer and Adder models.

While recent studies have either claimed \cite{santiDhar2013, iyerSrividya2014, priestman2017} or assumed \cite{campos2014, taheriJun2015, fievetDucret2015} that many bacterial species grow in size exponentially, some of them indicate linear growth at the level of a single cell \cite{messelink2021, chungKarAmir2024}. Since the determination of type of single-cell growth from the available experimental data is not always possible due to insufficient precision, researchers have developed indirect ways to differentiate between them \cite{karAmir2021, chungKarAmir2024}. Here we show that some of the above obtained single-cell-level distributions, statistical relationships, and correlations can also be used to differentiate between the linear and exponential growth regimes. Importantly, by comparing with published experimental data \cite{tanouchi2017, chungKarAmir2024}, we then show that these analytical predictions can be used to identify the underlying division and growth rules that a species may be following. Additionally, the growth rate of a cell might be correlated with the growth rate of its ancestors in the lineage \cite{taheriJun2015, tanouchi2017}. Therefore, we extend our analysis to include these autocorrelations and establish that our results are robust to such autocorrelations.


\section{Materials and Methods} \label{sec: materialMethods}
In this study, we obtain analytical probability distributions for various quantities related to bacterial cells in a culture, such as birth-size ($s_b)$, division-size $(s_d)$, division-time $(\tau_d)$, and division-added-size $(\Delta_d)$, where individual cells in the culture follow either the Timer, Sizer or Adder division rule and exhibit either linear or exponential growth. These quantities are called single-cell quantities because they concern characteristics of individual cells that do not evolve over their lifetime, as opposed to quantities like size and age, which evolve from birth to division. These are primarily obtained by tracking cells over multiple generations in a lineage, in microfluidic single-cell setups, such as the Mother Machine \cite{wangJun2010, taheriJun2015}. Consequently, their probability distributions are termed single-cell-level distributions. These single-cell quantities can, in principle, be also measured across a population in a batch type culture. However, recent studies have shown that the probability distributions for single-cell quantities sampled from lineage differ from those sampled from the bulk population \cite{genthon2019, genthon2020, genthon2022}. They are also more readily obtainable in mother-machine set-up as compared to batch type culture. Therefore, we restrict our analysis to the case where all of these quantities are sampled from the lineage.

In order to obtain these distributions, we start from three free parameters in our model -- the principal quantity, the growth rate, and the partitioning fraction. We refer to the distributions of these quantities across different individuals in a lineage as free-parameter distributions. Every division rule is defined by a principal single-cell quantity that determines when the division occurs. For the Timer model, this is the division-time, $\tau_d$ -- once the cell reaches this age, it divides. For the Sizer model, it is the division-size, $s_d$ -- the critical size at which the cell divides. For the Adder model, it is the added size at division, $\Delta_d$ -- the amount of size (mass or volume) a cell must add before dividing. The distribution corresponding to the principal quantity is what we call `principal distribution' for that division model. Besides indicating the division rule, the principal distribution also quantifies the rule's stochastic variability. For the Timer model, division-time distribution, $\Gamma(\tau_d)$, is the principal distribution. Similarly, for the Sizer and Adder models, division-size distribution, $\Xi(s_d)$, and division-added-size distribution, $\Omega(\Delta_d)$, are principal distributions respectively.

The second free-parameter distribution in our model is the growth-rate distribution, $\Lambda(\alpha)$, where the growth rates $\alpha$ corresponding to each cell in the lineage are sampled. Note that the single-cell-growth data from microfluidic experiments can be fitted as either linear or exponential growth. Therefore, two separate distributions are obtained for exponential and linear growth for the same data. The third distribution which is required for this analysis is partitioning-fraction distribution, $\kappa(\beta)$ -- the distribution of partitioning fraction, $\beta$, which is the fraction of size (mass or volume) that is inherited by the daughter cell from the parent cell. Since all single-cell quantities are related through relationships, such as $s_b = \beta s_d$, $\Delta_d = s_d - s_b$, $\tau_d = \Delta_d/\alpha$, and $s_d = s_b \exp(\alpha \tau_d)$, the remaining single-cell-level distributions can be obtained by statistical arguments and probability transformations, which are discussed in Appendix \ref{sec: SbirthSize} and \ref{sec: Sdistributions}.

Besides the closed-form analytical expressions for the distributions of single-cell quantities, analytical expressions for their means and standard deviations are also valuable, because they are measured more reliably in experiments than their distributions. Jun et al. \cite{taheriJun2015} derived some of these expressions for the Adder model under exponential growth for symmetric, non-stochastic partitioning, and Campos et al. \cite{campos2014} derived an expression for variance in birth-size in terms of principal distribution for the Adder model with asymmetric, stochastic partitioning. However, a comprehensive derivation of these statistical features of single-cell quantities for various growth and division models is lacking. Here, we derive the expressions for mean and standard deviation of these quantities for the Sizer and Adder models, exhibiting exponential or linear growth, with stochastic, asymmetric partitioning. Using these expressions, we further obtain relationships between the means of these quantities, between the standard deviations of these quantities, and correlations among these quantities (the derivations are detailed in Appendix \ref{sec: Srelationships} and \ref{sec: Scorrelations}). It turns out that some of these expressions and correlations can also be used to distinguish among these division and growth strategies.

It is also to be noted that the Timer model is left from our analysis because under exponential growth, as discussed earlier, it does not give cell-size homeostasis. Thus, the model can not be considered biologically realistic. Additionally, under the linear growth regime, the Timer model becomes equivalent to the Adder model. Both of them exhibit the same correlations, and the choice of $\Gamma(\tau_d)$ or $\Omega(\Delta_d)$ as the principal distribution does not alter the conclusions. This is further discussed in Appendix \ref{sec: StimerLinear}.

\section{Results} \label{sec: results}

\subsection{Single-cell-level distributions can differentiate among cell-division and growth models}  \label{subsec: analyticalDisbns}
\begin{figure}[t!]
    \centering
    \includegraphics[width=1\linewidth]{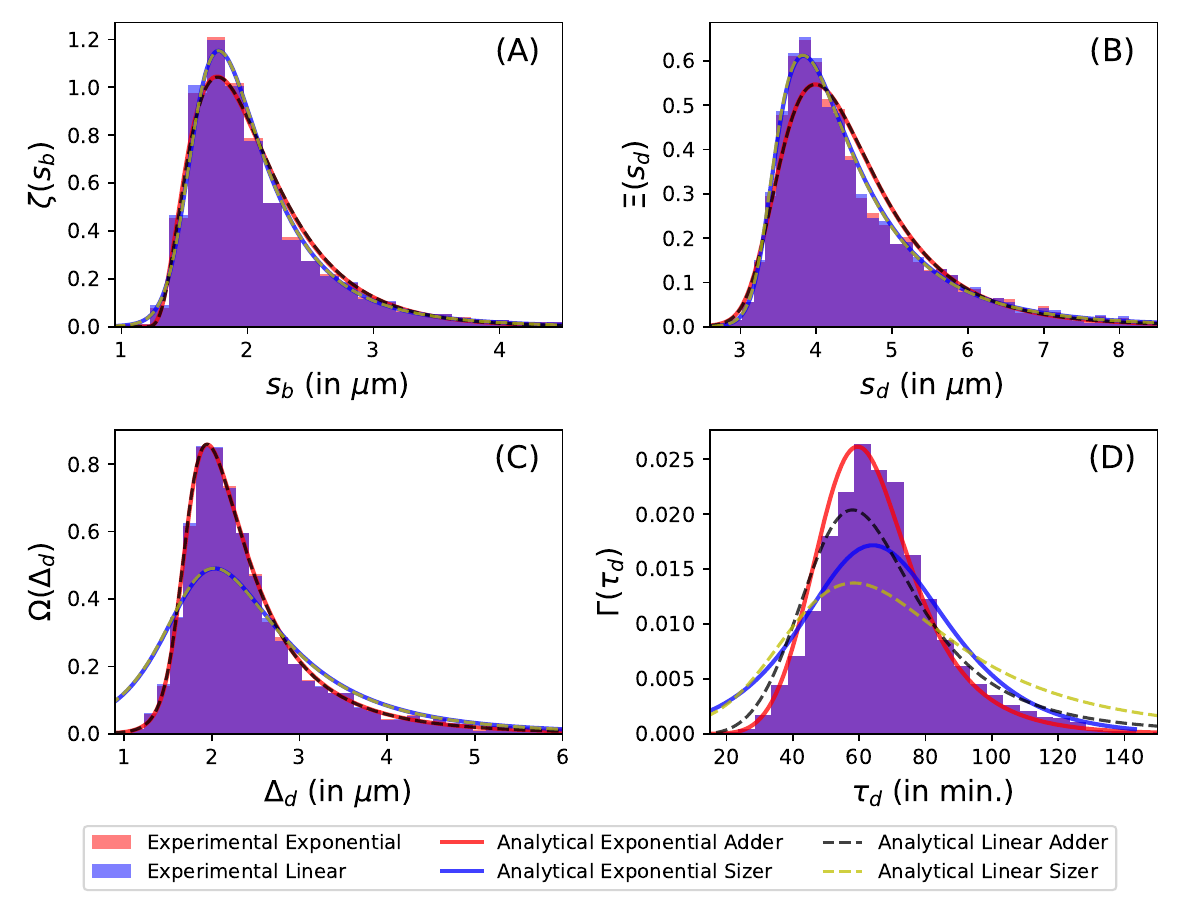}
\caption{Various experimental single-cell-level distributions (obtained from the single-cell experimental data for \textit{E. coli} ($25^{\circ}\text{C})$ \cite{tanouchi2017}) compared with their analytical results -- \textbf{(A)} Birth-size distribution, \textbf{(B)} Division-size distribution, \textbf{(C)} Division-added-size distribution, and \textbf{(D)} Division-time distribution, assuming a specific cell-division model (Sizer or Adder) and specific single-cell growth paradigm (linear or exponential). The three free-parameter distributions, namely the principal distribution ($\Xi(s_d)$ for the Sizer and $\Omega(\Delta_d)$ for the Adder model), growth-rate distribution $\Lambda(\alpha)$, and partitioning-fraction distribution $\kappa(\beta)$, are taken as Johnson SU fits to the corresponding experimental distributions. The remaining analytical distributions are obtained by statistical arguments and probability transformations. The two different histograms for linear and exponential growth, obtained from the same experimental data, arise because of the procedure used to improve the precision of division related quantities. This is done by extending the cell-size time-series (assuming either linear or exponential growth) to the time exactly in the middle of the two events of recording the last size for parent cell and the first size for daughter cell, which affects the estimated values of $s_b$, $s_d$, and $\Delta_d$ (explained further in Appendix \ref{subsec: SextraResultsSeparateHistograms}).}
    \label{fig:tanouchiDisbns25C}
\end{figure}

As discussed in Section \ref{sec: materialMethods}, there are three free-parameter distributions in our model: the principal distribution, i.e. $\Xi(s_d)$ for the Sizer model and $\Omega(\Delta_d)$ for the Adder model, the growth-rate distribution $\Lambda(\alpha)$, and the partitioning-fraction distribution $\kappa(\beta)$. Starting from these three distributions, one can obtain all other remaining single-cell-level distributions analytically (Appendix \ref{sec: SbirthSize} and \ref{sec: Sdistributions}). We further obtained various single-cell-level distributions for \textit{E. coli} and \textit{M. tuberculosis} from the data reported in two previous microfluidic studies \cite{tanouchi2017, chungKarAmir2024}. We then fitted Johnson family distributions to the these experimentally obtained distributions required for our analysis, such as  $\Xi(s_d)$, $\Omega(\Delta_d)$, $\Lambda(\alpha)$, and $\kappa(\beta)$, and used those resulting fits as the free-parameter distributions. Assuming either the Sizer or Adder division model and either linear or exponential single-cell growth, we subsequently derived all of the remaining distributions analytically. We then compared the analytical distributions with the corresponding experimental distributions to determine which of these matched the experimental data best, and thereby infer the division and growth model that most accurately describes the observed behavior of the bacterial species.

One such comparison for \textit{E. coli} at $25^{\circ}\text{C}$ is shown in Fig. \ref{fig:tanouchiDisbns25C}, where experimental single-cell-level distributions (taken from \cite{tanouchi2017}) are compared with the predicted analytical distributions for various division and growth models. It can be inferred from the figure that the analytical prediction corresponding to the exponential Adder model exhibits the closest agreement with the experimental data. This suggests that the microbial population (\textit{E. coli}) studied under these conditions ($25^{\circ}\text{C}$) follows a division strategy that is closer to the Adder model than to the Sizer model and exhibits single-cell growth that is closer to exponential than is linear in time. Note that to test the Sizer model, we took the best fitting $\Xi(s_d)$ distribution to be the principal distribution, and to test the Adder model, we took the best fitting $\Omega(\Delta_d)$ distribution as the principal distribution. 


Since Tanouchi et al. \cite{tanouchi2017} performed this experiment for the same strain of \textit{E. coli} at two additional temperatures ($27^{\circ}\text{C}$ and $37^{\circ}\text{C}$), we extended our analysis to these datasets also (refer to Fig. \ref{fig:tanouchiDisbns27C} and \ref{fig:tanouchiDisbns37C}). Although the comparison plots for these temperatures are less clear and the agreement is not as strong as obtained for the temperature $25^{\circ}\text{C}$, the experimental data are still most consistent with the exponential Adder model. In addition, we performed the same analysis on the experimental data obtained from another study for \textit{M. tuberculosis}  \cite{chungKarAmir2024} (refer to Fig. \ref{fig:karDisbnsSSBGFP}). Owing to the small number of data points, the experimental distributions obtained for various quantities showed substantial fluctuations. Likely due to this, the agreement between the predicted analytical and experimental distributions is not as close as that observed in Fig. \ref{fig:tanouchiDisbns25C}. Furthermore, it can be inferred from Fig. \ref{fig:karDisbnsSSBGFP} that the choice between linear and exponential growth has only a minor effect on the distributions. Despite these discrepancies, the data suggest the Adder model is a better candidate for the division model than the Sizer model.

It is to be noted that the experimental distributions for these single-cell quantities exhibit long-tail behavior. Therefore, in order to obtain the free-parameter distributions, we also examined several alternative distributions for the purpose of fitting them to the required experimental distributions. These include  gamma, Gumbel, and log-normal distributions. However, the Johnson family of distributions consistently provided the best fit to the data. Additionally, the analytical distributions of the other derived quantities matched the experimental distributions more closely when the free-parameter distributions were taken to be the best-fitting Johnson-family distributions instead of gamma, Gumbel, or log-normal distributions (Fig. \ref{fig:otherFits}). Hence, the Johnson family of distributions were chosen to fit the experimental distributions.

\subsection{Statistical relationships between single-cell quantities can distinguish among cell-division and growth models} \label{subsec: relationships}
\begin{figure}[h!]
    \centering
    \includegraphics[width=0.8\linewidth]{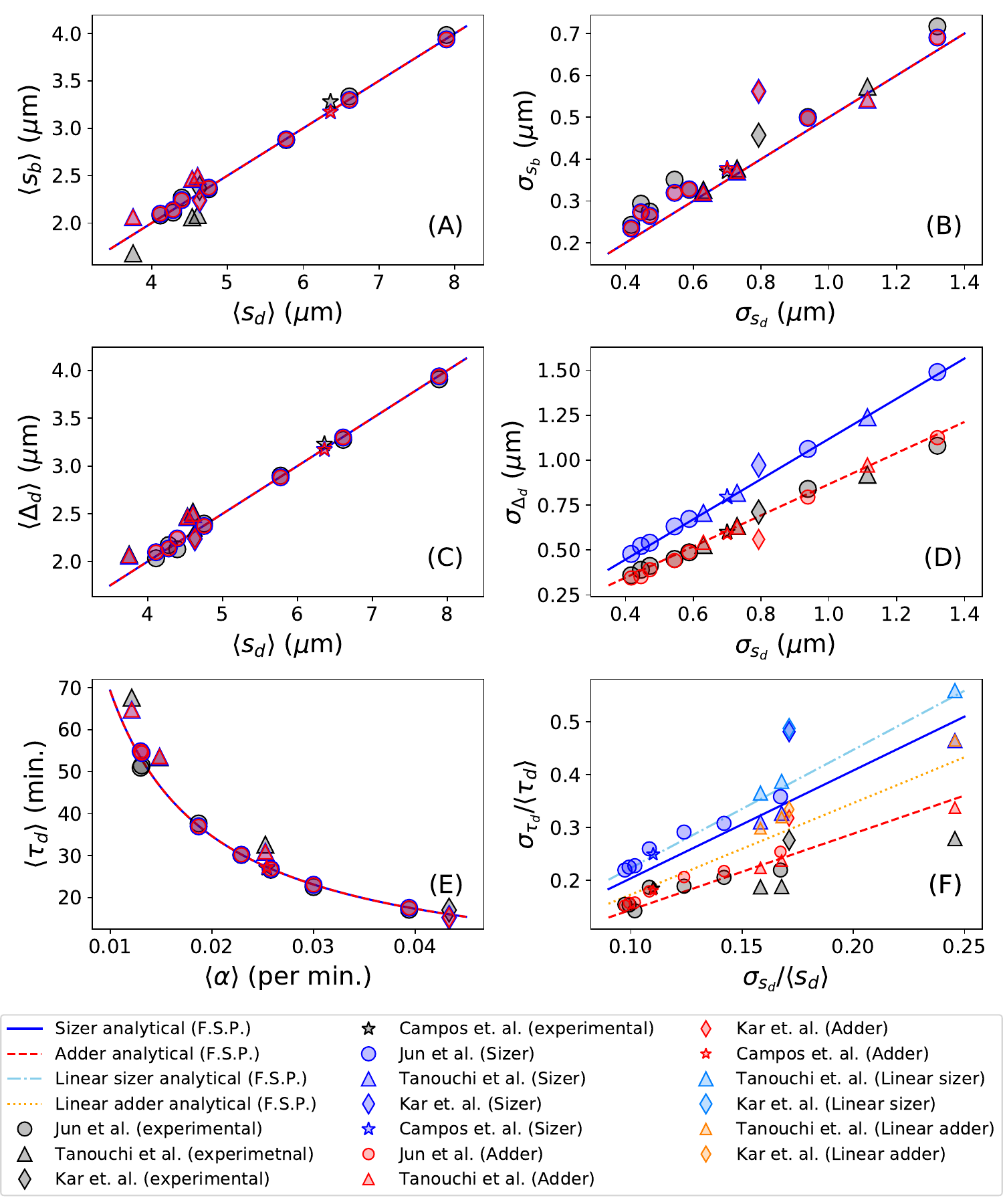}
    \caption{Mean, standard deviation, and coefficient of variation (the ratio of standard deviation and mean) for various single-cell quantities plotted against each other -- \textbf{(A)} mean birth-size versus mean division-size, \textbf{(B)} standard deviation of birth-size versus standard deviation of division-size, \textbf{(C)} mean division-added-size versus mean division-size, \textbf{(D)} standard deviation of division-added-size versus standard deviation of division-size, \textbf{(E)} mean division-time versus mean exponential growth rate, \textbf{(F)} coefficient of variation of division-time versus coefficient of variation of division-size. The data have been taken from previous studies \cite{tanouchi2017, taheriJun2015, campos2014, chungKarAmir2024}. For Tanouchi et al. \cite{tanouchi2017}, the data points corresponding to \textit{E. coli} at all three temperatures investigated in the study are plotted. For Jun et al. \cite{taheriJun2015}, the data points corresponding to \textit{E. coli} in all seven growth media studied are plotted. For Kar et al. \cite{chungKarAmir2024}, the data points corresponding to \textit{M. tuberculosis} strain CDC1551 labeled with SSB-GFP are plotted. For Campos et al. \cite{campos2014}, the data points corresponding to \textit{E. coli} strain BW25113 in LB rich medium are plotted. The symbols with black color correspond to the experimental data from different studies and the same symbol with other colors correspond to the corresponding analytical predictions for different division and growth models. \textbf{F.S.P.} in the legend refers to ``\textbf{fixed symmetric partitioning}'', which means that cell division is symmetric and there is no stochasticity in partitioning. The points corresponding to the Sizer and Adder predictions for various studies in \textbf{(A)}, \textbf{(B)}, \textbf{(C)}, and \textbf{(E)} lie on top of each other. Only in panel \textbf{(F)}, extra points are plotted for linear Sizer and linear Adder, and the points labeled `Sizer' and `Adder' correspond to exponential Sizer and exponential Adder models, respectively. Additionally, only for this panel, the blue line and red dashed line correspond to the predictions of exponential Sizer and exponential Adder for fixed stochastic partitioning.}
    \label{fig: statisticalRelationships}
\end{figure}
As mentioned earlier, we obtained relationships between the means of single-cell quantities and between their standard deviations. These relationships are derived in Appendix \ref{sec: Srelationships}. Some relationships between the means of single-cell quantities are universal across growth and division models, whereas others depend on the specific growth model. However, none of these mean-based relationships uniquely identifies the underlying growth or division mechanism, as discussed further below. In contrast, the relationships between the standard deviations of certain quantities can distinguish between the Adder and Sizer division strategies. Interestingly, a subset of these relationships can further differentiate between single-cell growth types.

For the case of asymmetric, stochastic partitioning, the relationship between the averages of size-related single-cell quantities can be written as:
\begin{equation} \label{eq: meanSbSdDeltaD}
    \langle s_b \rangle \;=\;  \langle \beta \rangle \langle s_d \rangle \;=\; \frac{\langle \beta \rangle}{1-\langle \beta \rangle} \langle \Delta_d \rangle
\end{equation}
Also, for both the Adder and Sizer models, the mean division-time $\langle \tau_d \rangle$ under exponential single-cell growth is given by:
\begin{equation} \label{eq: meanTauDexponential}
    \langle \tau_d \rangle = \frac{-\ln{\langle \beta \rangle}}{\langle \alpha_E \rangle}
\end{equation}
where $\langle \alpha_E \rangle$ is average exponential growth rate, and $\langle \beta \rangle$ is the average partitioning fraction. Similarly, for the case of linear growth, irrespective of the Adder or Sizer model, one can write for $\langle \tau_d \rangle$:
\begin{equation} \label{eq: meanTauDlinear}
    \langle \tau_d \rangle \;\;= \frac{\langle \Delta_d \rangle}{\langle \alpha_L \rangle} \;\;= \frac{(1-\langle \beta \rangle)\langle s_b \rangle}{\langle \beta \rangle \langle \alpha_L \rangle} \;\;= \frac{(1-\langle \beta \rangle) \langle s_d \rangle}{\langle \alpha_L \rangle}
\end{equation}
where $\langle \alpha_L \rangle$ is the average linear growth rate. It may appear that the two equations above can differentiate between linear and exponential growth types. However, the $\langle \alpha_E \rangle$ and $\langle \alpha_L \rangle$ obtained by fitting the same experimental data with exponential or linear  growth will lead to the same $\langle \tau_d \rangle$ when used in Eq. \ref{eq: meanTauDexponential} and \ref{eq: meanTauDlinear} respectively. Therefore, we see that none of the relationships between the averages of single-cell quantities can distinguish between these division models. In Fig. \ref{fig: statisticalRelationships}, mean, standard deviation, and coefficient of variation of various single-cell quantities are plotted against each other for several bacterial species using the data from previous studies \cite{tanouchi2017, taheriJun2015, chungKarAmir2024, campos2014}. Along with them, the predictions of different division (the Sizer and Adder) and growth models (linear and exponential) for these species are also shown as different symbols with various colors, whereas the symbols with black color are associated with the experimental data. The predictions for these division and growth models reduce to simpler relationships for the case of fixed symmetric partitioning (see Appendix \ref{sec: Srelationships}), where binary division takes place and each daughter cell gets exactly half the volume of the parent cell at division without any stochasticity. Therefore, along with the symbols, different lines corresponding to different division and growth models are shown for the case of fixed symmetric partitioning. Specifically, the panels (A), (C), and (E) of Fig. \ref{fig: statisticalRelationships} present comparisons of mean values of the single-cell quantities. Apart from being very close to the experimental data points, the symbols corresponding to the analytical predictions for the Sizer and Adder models lie on top of each other. This illustrates that, although the relationships between the mean values of these quantities are in excellent agreement with the experimental data, they are unable to differentiate between these division models.

Additionally, some other relationships between the standard deviations of single-cell quantities are also invariant to the division and growth models. For example, the relationship between the standard deviation of $s_d$ and standard deviation of $s_b$ is given as:
\begin{equation} \label{eq: sigmaSbSigmaSd}
    \sigma_{s_b} \;=\; \langle \beta \rangle \sigma_{s_d} \sqrt{ 1 + \left( \frac{\sigma_\beta}{\langle \beta \rangle} \right)^2 + \left( \frac{C(\beta)}{C(s_d)} \right)^2}
\end{equation}
where $C(x) = \sigma_x/\langle x \rangle$ is the coefficient of variation of $x$. This relationship holds for both the Sizer and Adder models under linear as well as exponential growth, and therefore can not be used to distinguish between these models, as shown in Fig. \ref{fig: statisticalRelationships} (B).

In contrast to the relationships that are true for both the Sizer and Adder models, there are some relationships between the standard deviations of size-related single-cell quantities, such as $s_b$, $s_d$, and $\Delta_d$, which can differentiate between these two division models. One such example is the relationship between $\sigma_{\Delta_d}$ and $\sigma_{s_d}$. For the Sizer model, this relationship is given as:
\begin{equation} \label{eq: sigmaDeltaDsdSizer}
    \sigma_{\Delta_d} \;=\;  \sigma_{s_d} \sqrt{ 1 + \langle \beta \rangle^2 \left( 1 + \left( \frac{C(\beta)}{C(s_d)} \right)^2 \right) + \sigma_{\beta}^2}.
\end{equation}
Whereas, for the Adder model, $\sigma_{\Delta_d}$ and $\sigma_{s_d}$ are related as:
\begin{equation} \label{eq: sigmaDeltaDsdAdder}
    \sigma_{\Delta_d} \;=\;  \sigma_{s_d} \sqrt{ 1 - \langle \beta \rangle^2 \left( 1 + \left( \frac{C(\beta)}{C(s_d)} \right)^2 \right) - \sigma_{\beta}^2}.
\end{equation}
From Fig. \ref{fig: statisticalRelationships} (D), it is apparent that this relationship can be used to differentiate between the Adder and Sizer models, where the experimental data points are closer to the symbols corresponding to the Adder model than to those of the Sizer model. Therefore, it can be concluded that the bacterial species represented by various symbols in this plot appear to follow a division strategy that is closer to the Adder model than to the Sizer model.

Note that, the relationships mentioned so far are independent of the growth type. Therefore, in Fig. \ref{fig: statisticalRelationships} (A-E), the symbols corresponding to the predictions for the Sizer and Adder models are plotted without specifying the growth type. In contrast, the relationship between $\sigma_{\tau_d}$ and $\sigma_{s_d}$ differs between the Sizer and Adder models as well as between linear and exponential growth types. When the single-cell growth is exponential, one obtains the following expression for the Sizer model:
\begin{equation} \label{eq: COVtauDexpSizer}
    \frac{\sigma_{\tau_d}}{\langle \tau_d \rangle} \;=\; \sqrt{ \frac{1}{(\ln{\langle \beta \rangle})^2}  \left[  C(s_d)^2 + C(s_b)^2 \right] + C(\alpha_E)^2 }.
\end{equation}
Whereas for the Adder model, this relationship is slightly different:
\begin{equation} \label{eq: COVtauDexpAdder}
    \frac{\sigma_{\tau_d}}{\langle \tau_d \rangle} \;=\; \sqrt{ \frac{1}{(\ln{\langle \beta \rangle})^2}  \left[  C(s_d)^2 + (1-2\langle \beta \rangle) C(s_b)^2 \right] + C(\alpha_E)^2 }.
\end{equation}
In the equations above, $\alpha_E$ is the exponential growth rate. Similarly, when the single-cell growth is linear, one can write the following expression for the Sizer model:
\begin{equation} \label{eq: COVtauDlinearSizer}
    \frac{\sigma_{\tau_d}}{\langle \tau_d \rangle} \;=\; \sqrt{ \frac{C(s_d)^2 + \langle \beta \rangle^2 C(s_b)^2}{(1-\langle \beta \rangle)^2}  + C(\alpha_L)^2 }.
\end{equation}
In contrast, for the Adder model, this relationship is given as:
\begin{equation} \label{eq: COVtauDlinearAdder}
    \frac{\sigma_{\tau_d}}{\langle \tau_d \rangle} \;=\; \sqrt{ \frac{C(s_d)^2 - \langle \beta \rangle^2 C(s_b)^2}{(1-\langle \beta \rangle)^2}  + C(\alpha_L)^2 }
\end{equation}
Here $\alpha_L$ is linear growth rate. Further using Eq. \ref{eq: meanSbSdDeltaD} and \ref{eq: sigmaSbSigmaSd}, one can eliminate $\langle s_b \rangle$ and $\sigma_{s_b}$ from Eq. \ref{eq: COVtauDexpSizer}, \ref{eq: COVtauDexpAdder}, \ref{eq: COVtauDlinearSizer}, and \ref{eq: COVtauDlinearAdder}, and obtain these relationships solely in terms of $\tau_d$, $s_d$, and $\alpha_E$ or $\alpha_L$. As evident from the equations above (Eq. \ref{eq: COVtauDexpSizer}, \ref{eq: COVtauDexpAdder}, \ref{eq: COVtauDlinearSizer}, and \ref{eq: COVtauDlinearAdder}), the relationship between coefficient of variation of $\tau_d$ and coefficient of variation of $s_d$ can differentiate between the Sizer and Adder models, as well as between linear and exponential growth types (Fig. \ref{fig: statisticalRelationships} (F)). Note that, in this sub-figure, the symbols labeled `Adder' and `Sizer' correspond to exponential Adder and exponential Sizer models, respectively. The symbols labeled `Linear Adder' and `Linear Sizer' have usual meanings. However, they are shown only for the datasets of Tanouchi et al. \cite{tanouchi2017} and Kar et al.\cite{chungKarAmir2024} because the estimation of these points requires the coefficient of variation of linear growth rate (see Eq. \ref{eq: COVtauDlinearSizer} and \ref{eq: COVtauDlinearAdder}), which is only available from these two studies.

As shown in the Fig. \ref{fig: statisticalRelationships} (F), for the three temperatures reported by Tanouchi et al. (triangles), the experimental results show closer agreement with the predictions of the exponential Adder model than with those of the other models. This suggests that, at all the three temperatures, the bacterial species follows a division strategy that is more consistent with the Adder model than with the Sizer model, while its single-cell growth dynamics are more consistent with exponential growth than with linear growth. The same conclusion was drawn from the analysis done in Section \ref{subsec: analyticalDisbns}. Similarly, among the data points corresponding to Kar et al. (diamonds), the symbol representing experimental data is very close to the symbols corresponding to both linear Adder and exponential Adder. Therefore, also concluded in Section \ref{subsec: analyticalDisbns}, the type of single-cell growth can not be differentiated between linear and exponential for this case.

\subsection{Correlations between single-cell quantities can differentiate among cell-division and growth models} \label{subsec: correlations}

\begin{figure}[h!]
    \centering
    \includegraphics[width=1\linewidth]{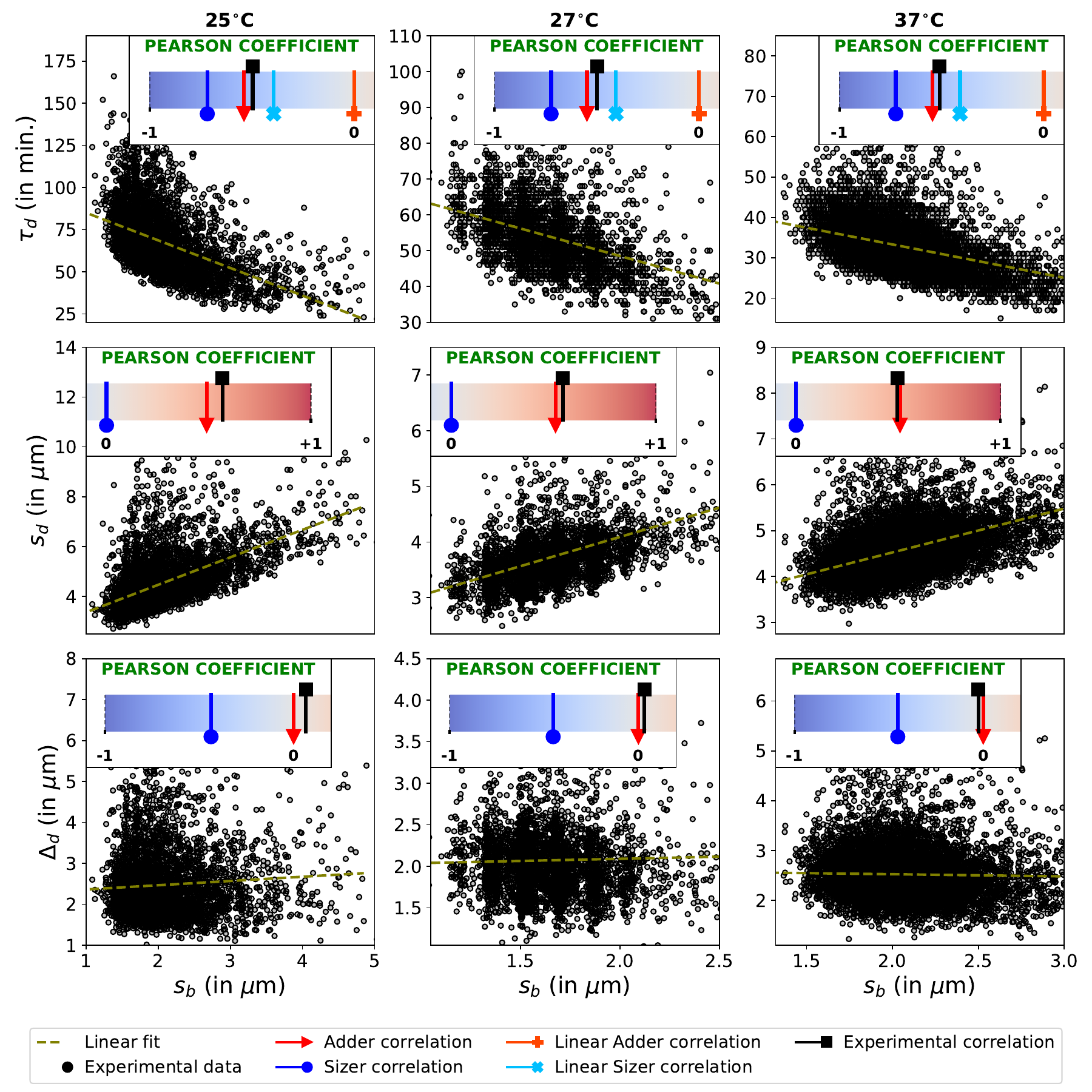}
    \caption{Various single-cell quantities, such as $\tau_d$, $s_d$, and $\Delta_d$, plotted against $s_b$ to infer correlations of these quantities with birth-size. The experimental data have been taken from a previous study \cite{tanouchi2017}  for \textit{E. coli} at different temperatures --  $25^{\circ}\text{C}$, $27^{\circ}\text{C}$, and $37^{\circ}\text{C}$. In the first row, $\tau_d$ is plotted against $s_b$; in the second row, $s_d$ is plotted against $s_b$; and in the third row, $\Delta_d$ is plotted against $s_b$. The first column represents these correlations at $25^{\circ}\text{C}$, the second column represents the correlations at $27^{\circ}\text{C}$, and the third column represents the correlations at $37^{\circ}\text{C}$. The
dashed line in the scatter plot represents the linear fit to the data. The insets in all of these plots compare `Pearson Correlation Coefficients' for the experimental data with its predicted analytical values for various growth and division models as indicated in the legend. For the insets corresponding to the correlation between $\tau_d$ and $s_b$ (first row), the labels corresponding to Adder and Sizer correlations (red triangle with a line and blue circle with a line) refer to exponential Adder and exponential Sizer, respectively. For other rows, the correlation values are independent of growth type and hence only symbols/labels corresponding to Adder and Sizer are shown.}
    \label{fig:correlations}
\end{figure}

\begin{table}[h!]
    \centering
    \renewcommand{\arraystretch}{2}
    \begin{tabular}{|c||c|c|c|c|}
        \hline
        \textbf{Correlation}& \textcolor{blue}{\textbf{Exponential Sizer}}& \textcolor{deepskyblue}{\textbf{Linear Sizer}}& \textcolor{red}{\textbf{Exponential Adder}}& \textcolor{OrangeRed}{\textbf{Linear Adder}}\\
        \hline
        \hline
        $C(\tau_d,s_b)$ & $-\sigma_{s_b}/(\langle \alpha_E \rangle \langle s_b \rangle \sigma_{\tau_d})$& $-\sigma_{s_b}/(\langle \alpha_L \rangle \sigma_{\tau_d})$& $\sigma_{s_b}(\langle \beta \rangle -1)/(\langle \alpha_E \rangle \langle s_b \rangle \sigma_{\tau_d})$& 0\\
        \hline
        $C(s_d,s_b)$ & 0 & 0 & $\sigma_{s_b}/\sigma_{s_d}$& $\sigma_{s_b}/\sigma_{s_d}$\\
        \hline
        $C(\Delta_d,s_b)$ & $-\sigma_{s_b}/\sigma_{\Delta_d}$ & $-\sigma_{s_b}/\sigma_{\Delta_d}$ & 0 & 0\\
        \hline
    \end{tabular}
    \caption{Various correlations between different single-cell quantities for different division and growth models are summarized here. In the expressions for the correlations, $\alpha_E$ and $\alpha_L$ correspond to exponential and linear growth rates, respectively. Different rows represent correlations between different quantities, whereas different columns represent different division and growth models for which the correlations are shown. The first row shows the correlation between division-time and birth-size. Similarly, second and third rows depict the correlation of division-size with birth-size, and division-added-size with birth-size, respectively. The first column shows the correlations for exponential Sizer model. Similarly, the subsequent columns represent the correlations for linear Sizer, exponential Adder, and linear Adder model, respectively.}
    \label{table}
\end{table}

As discussed in the Introduction, recent studies widely use correlations between $\Delta_d$ and $s_b$, between $s_d$ and $s_b$, and  between $\tau_d$ and $s_b$ to differentiate among the various cell-division models, such as the Timer, Sizer and Adder. These correlations are derived for the Adder model under exponential growth and symmetric, non-stochastic partitioning \cite{taheriJun2015}. But the cell partitioning at division is inherently stochastic and can be slightly asymmetric. Therefore, it is necessary to assess how asymmetry and stochasticity affect these correlations. Here we derive the correlations for asymmetric, stochastic partitioning for the Sizer and Adder models with single-cell growth being either linear, or exponential. Thereby, we show that these correlations can discriminate not only among division mechanisms, but also among single-cell-growth types. These correlations are summarized in Table \ref{table}, and they are derived in Appendix \ref{sec: Scorrelations}.

It can be seen from Table \ref{table} that the correlation between size-related quantities, such as between $s_d$ and $s_b$, and between $\Delta_d$ and $s_b$, differ between the Sizer and Adder models. Consequently, these correlations can be used to determine which division-control model is followed by the bacterial species under study. In contrast, because these correlations are independent of the single-cell growth, they can not be used to infer the growth type. Instead, one has to use correlation between $\tau_d$ and $s_b$ that depends upon single-cell-growth type as well as the division model followed by the bacterial species (which can also be seen from Table \ref{table}). Hence, after determining the division model from the correlations between size-related quantities, one can further use the correlation between $\tau_d$ and $s_b$ to differentiate between linear and exponential growth.

In Fig. \ref{fig:correlations}, several experimentally measured single-cell quantities, such as $\tau_d$, $s_d$, and $\Delta_d$, are plotted against $s_b$ to infer these correlations at three different temperatures for which the bacterial species is studied \cite{tanouchi2017}. It is evident from the size-related correlations (between $s_d$ and $s_b$, and between $\Delta_d$ and $s_b$) that the division model followed by the bacterial species across all temperatures is close to the Adder model. This can be seen from the Pearson coefficient value of the experimental data. Additionally, from the correlation between $\tau_d$ and $s_b$, we can further observe that the correlation for the experimental data is in the best agreement with the prediction of exponential Adder. Therefore, it may be concluded that the bacterial species studied in this experiment exhibits a division strategy that is most consistent with the Adder model, together with nearly exponential single-cell growth, at all three temperatures reported. Note that the same conclusions were drawn from Section \ref{subsec: analyticalDisbns} and \ref{subsec: relationships}.

A similar analysis was also done for the data obtained from \cite{chungKarAmir2024}  (refer to Fig. \ref{fig: correlationsKar}). As concluded in the Section \ref{subsec: analyticalDisbns} and \ref{subsec: relationships}, this analysis also indicates that the bacterial species under study follows the Adder division model, while the type of single-cell growth can not be discriminated reliably between linear and exponential growth types.

\subsection{The results are robust to autocorrelations in single-cell growth rate} \label{subsec: autocorrelations}
The three free parameters in our model -- the principal quantity ($s_d$ for the Sizer model and $\Delta_d$ for the Adder model), $\alpha$, and $\beta$, are assumed to be independent of each other. To check if this assumption holds true in reality, we analyzed the single-cell data from Tanouchi et al. \cite{tanouchi2017} to infer correlations between them. The correlations between partitioning fraction ($\beta$) and other quantities ($s_b$, $s_d$, $\Delta_d$, $\tau_d$, and $\alpha$) were found to be negligible as were the correlations between the growth rate ($\alpha$) and other quantities ($s_d$, $s_b$, $\Delta_d$, and $\beta$) (Pearson Correlation Coefficient $\leq$ 0.1). These correlations are shown in Fig. \ref{fig:crossCorrelations}. Likewise, no significant correlations were found between $\beta$ and its values in preceding generations within the lineage. However, the growth rate of a cell exhibited significant correlations with the growth rates of cells in preceding generations of the same lineage (Fig. \ref{fig:autoCorrelations}), consistent with observations reported by \cite{taheriJun2015}.

Although the data shows that growth rate of a cell is correlated with the growth rate of its ancestor cells, our simulations show that these autocorrelations in growth rate do not affect the single-cell-level distributions obtained analytically for the case when such correlations are absent (see Fig. \ref{fig:autoCorrelationsAlpha}). The reason is that the size-related distributions, such as $\zeta(s_b)$, $\Xi(s_d)$, and $\Omega(\Delta_d)$, are obtained only from the rules of division and partitioning (principal distribution and partitioning-fraction distribution). Therefore, the single-cell-growth type and single-cell-growth rate do not affect the analytical forms of these distributions. In contrast, the remaining distribution, $\Gamma(\tau_d)$, is obtained using all of the three free-parameter distributions, which include $\Lambda(\alpha)$. But given that the growth rate autocorrelations do not affect $\Lambda(\alpha)$, $\Gamma(\tau_d)$ will also remain the same.

Further, since these autocorrelations have no effect on the full distribution of these quantities, their means and standard deviations also remain unchanged. This makes our statistical relationships regarding single-cell quantities robust to such autocorrelations. Furthermore, the correlations between single-cell quantities within the same generation are also independent of these autocorrelations because they are derived only from the fundamental correlations, such as the correlation between principal quantity and $s_b$. Therefore, the results mentioned previously in Section \ref{subsec: analyticalDisbns}, \ref{subsec: relationships}, and \ref{subsec: correlations} are robust to such autocorrelations in the growth rate.

The other autocorrelations, such as those for $s_d$, $s_b$, $\Delta_d$, and $\tau_d$, can also be calculated analytically and measured experimentally (Fig. \ref{fig:autoCorrelations}). However, it can be seen in the figure that these measured autocorrelations are not very reliable. Also, they do not provide significant new information than those already available through the above calculated cross-correlations between these quantities (Fig. \ref{fig:correlations}).

\section{Discussion} \label{sec: discussion}
Various cell-division strategies, like the Timer, Sizer, and Adder, have different rules regarding cell division. This study shows that these division, as well as growth strategies lead to observable differences in the probability distributions of various single-cell quantities, correlations among them, and the statistical relationships between these quantities. These differences can then be used to differentiate between the various growth and division models. We used experimental data from previous studies \cite{tanouchi2017, chungKarAmir2024, campos2014, taheriJun2015} to show that our predictions can indeed identify the underlying growth and division rules that a species may be following.


One aspect worth noting in our study is that, despite the overall good agreement between the experimental data and the analytical predictions, there exists a slight mismatch between them. Although the discrepancy may arise from insufficient experimental precision in measuring single-cell quantities, or an insufficient sample size, a possible alternative theoretical explanation also cannot be ruled out. Indeed, some studies have suggested mixed models for cell division and growth, such as bilinear growth \cite{donachie1976, kubitschek1981}, super-exponential growth \cite{heerden2023, cylkeBanerjee2023}, a mixed size and age-based division strategy for \textit{E. coli} \cite{osellaNugent2014}, biphasic growth for \textit{M. smegmatis} \cite{hannebelle2020}, and linear growth until the cell reaches a critical size, followed by an exponential growth for a constant time for \textit{B. subtilis} \cite{nordholtHeerden2020}. Therefore, further extension of present analysis to handle such scenarios could potentially improve agreement with experimental data and constitutes an open avenue for future investigation.

In recent years, many studies have focused on the intrinsic molecular details of cell-division models \cite{walldenLundis2016, siJun2019, pandeyJain2020, serbanescuBanerjee2020, nietoGarcia2024, nietoGarcia2024B}. Our study here adopts a coarse-grained approach and does not explicitly incorporate these molecular details. This makes our results robust to the details of molecular implementations of these growth and division strategies. Nevertheless, the search for molecular implementations remains an exciting endeavor.

\appendix
\renewcommand{\thefigure}{\thesection\arabic{figure}}
\makeatletter
\@addtoreset{figure}{section}
\makeatother

\section{Birth-size distributions} \label{sec: SbirthSize}

Birth-size distributions are critical in determining other single-cell-level distributions for the Sizer and Adder models. One obtains $\zeta(s_b)$ for a division model from the principal distribution and $\kappa(\beta)$,  and then obtains the other remaining distributions through probability transformations using relationships between the single-cell quantities. However, given that the individual cells grow exponentially, $\zeta(s_b)$ can not be obtained for the Timer model due to lack of cell-size homeostasis, as noted in the Introduction and discussed further below.
\subsection{Timer model} \label{subsec: SbirthSizeTimer}
Consider a Mother Machine setup in which individual cells are tracked over multiple generations. We assume that cells grow exponentially in size and divide symmetrically into two equal daughter cells without partitioning stochasticity, following the Timer model of cell division. The birth-size for a daughter cell is given by $s_{b_d} = s_b \exp{(\alpha \tau_d)}/2$, where $s_b$, $\alpha$, and $\tau_d$ are birth-size, growth rate, and division-time, respectively, for its parent cell. $\alpha$ and $\tau_d$ are treated as two independent quantities for the Timer model, and one can choose the desired values for both of them independently (the same is not true for the Sizer and Adder models, where the division-time is found from growth rate and other quantities using the relationship $\tau_d = \ln(s_d/s_b)/\alpha$). If $\exp{(\alpha \tau_d)} \neq 2$, the birth-size for the daughter cell will be either smaller or bigger than that of the parent cell. And therefore, it can not be guaranteed that the value of birth-size for the tracked cell will approach a constant value over generations. But, with properly chosen non-stochastic values of $\alpha$ and $\tau_d$ such that $\tau_d = \ln{(2)/\alpha}$ (along with no partitioning stochasticity), the two daughter cells created after cell-division have birth-sizes equal to that of the parent cell. Thus the birth-size remains constant over generations for the Timer model. However, adding stochasticity in these parameters destroys the size homeostasis.

To prove this, let us now assume that the division-time is a stochastic quantity and its probability distribution is a Gaussian distribution (mean $\langle \tau_d \rangle$ and standard deviation $\sigma_{\tau_d}$), while growth rate and partitioning fraction are still non-stochastic quantities ($\beta = 0.5$). The ratio of the birth-size of the daughter cell to that of the parent cell is given by $\lambda = \exp{(\alpha \tau_d)}/2$ because $s_{b_d} = s_b \exp{(\alpha \tau_d)}/2$. Hence, the distribution of $\lambda$, $\Upsilon(\lambda)$ is given by:
\begin{equation}
    \Upsilon(\lambda) = \frac{1}{\alpha \lambda\sqrt{2 \pi \sigma_{\tau_d}^2}} \exp{\left(  \frac{-(\frac{\ln{(2\lambda)}}{\alpha} - \langle \tau_d \rangle)^2}{2\sigma_{\tau_d}^2}\right)}.
\end{equation}
And, the average value of $\lambda$ will be given as:
\begin{equation}
    \langle \lambda \rangle \;=\; \int_{0}^{\infty} d\lambda \; \lambda \; \Upsilon(\lambda) \;=\; \frac{1}{\alpha \sqrt{2 \pi \sigma_{\tau_d}^2}}\int_{0}^{\infty} d\lambda \; \exp{\left(  \frac{-(\frac{\ln{(2\lambda)}}{\alpha} - \langle \tau_d \rangle)^2}{2\sigma_{\tau_d}^2}\right)}.
\end{equation}
Further doing the manipulation of $y=\ln{(2 \lambda)}$, one can get:
\begin{equation}
    \langle \lambda \rangle \;=\; \frac{1}{2\alpha \sqrt{2 \pi \sigma_{\tau_d}^2}}\int_{-\infty}^{\infty} d y \; \exp{\left(y -  \frac{(y-\alpha \langle \tau_d \rangle)^2}{2 \alpha^2 \sigma_{\tau_d}^2}\right)}.
\end{equation}
Using the method of completing the square, the integral above can be evaluated, and it can be further shown that
\begin{equation}
    \langle \lambda \rangle \;=\; \frac{e^{\alpha \langle \tau_d \rangle}}{2} \exp{\left(\frac{\alpha^2 \sigma_{\tau_d}^2}{2}\right)} .
\end{equation}
First of all, for the case of non-stochastic division-times ($\sigma_{\tau_d}=0$), the equation above shows that the values of $\alpha$ and $\langle \tau_d \rangle$ must be fine tuned in order to make $\langle \lambda \rangle =1$, which ensures that the birth-size for the daughter cell is equal to that for the parent cell on average. However, even for properly chosen value of $\alpha$ and $\langle \tau_d \rangle$, if there is stochasticity in $\tau_d$, the average value of $\lambda$ is greater than $1$ for any desired value of $\sigma_{\tau_d}$. This means that, although a daughter cell is equally likely to divide earlier or later than the average division-time $\langle \tau_d \rangle$ (because $\Gamma(\tau_d)$ is symmetric), its birth-size will be greater than that of its parent on average. This is so because the cells with bigger division-times will accumulate disproportionately more biomass in comparison to the cells with smaller division-times due to exponential behavior of the cell-size growth. This results in the average birth-size in the population increasing with time, similar to the one seen in the inset for the Timer model in Fig. \ref{fig: cartoon}. Similar calculations can be done for the case when the division-time and partitioning fraction are non-stochastic and the growth rate is stochastic with Gaussian distribution. That case also gives the same results. Asymmetry and stochasticity in partitioning further make it difficult for the Timer model to achieve cell-size homeostasis.

\subsection{Sizer and Adder models} \label{subsec: SbirthSizeSizerAdder}
For the Sizer and Adder models, we present some statistical arguments that are used to obtain $\zeta(s_b)$ from principal distributions. The first argument goes like this. The birth-size of a random cell in a lineage is given by $s_b = \beta s_d$. Here, $s_d$ is the size of the parent cell at the time of division, and is sampled from the probability distribution $\Xi(s_d)$. Also, $\beta$ is the the fraction of total size (mass or volume) of the parent cell that is inherited by the daughter cell at the time of division, and it is sampled from the probability distribution $\kappa(\beta)$. Additionally, $\beta$ is assumed to be independent of all other single-cell quantities including $s_d$ (discussed in Section \ref{subsec: autocorrelations}) . Therefore, the distribution for $s_b$ can be written in terms of the distributions for $s_d$ and $\beta$ as:
\begin{equation} \label{eq: zetaSizer}
    \zeta(s_b) \;=\; \int_{0}^{\infty} ds_d \; \frac{\Xi(s_d)}{s_d} \; \kappa \left(\frac{s_b}{s_d} \right) \;=\; \int_{0}^{\infty} d\beta \; \Xi \left( \frac{s_b}{\beta} \right) \;\frac{\kappa(\beta)}{\beta}.
\end{equation}
Since for the Sizer model, $\Xi(s_d)$ and $\kappa(\beta)$ are free-parameter distributions, $\zeta(s_b)$ is obtained using this expression. For the case of symmetric, non-stochastic partitioning, i.e. $\kappa(\beta) \;=\; \delta(\beta - 1/2)$, the expression for $\zeta(s_b)$ can be simplified as:
\begin{equation}
    \zeta(s_b) \;=\; 2 \;\Xi(2 s_b) .
\end{equation}
In order to obtain $\zeta(s_b)$ for the Adder model from the free-parameter distributions, namely $\Omega(\Delta_d)$ and $\kappa(\beta)$, we select a random cell in a cell lineage with birth-size $s_b$, which was created from a parent cell as a result of the cell division. This parent cell had a birth-size $s_1$, and it had added a certain size $\Delta_{d_1}$ since its birth to the time of its division ($\Delta_{d_1}$ is the division-added-size for the parent cell). Hence, one can write $s_b = \beta_1(s_{1} + \Delta_{d_1})$, where $\beta_1$ is the fraction of the size inherited by the daughter cell from the parent cell after division. Further tracking the parent cell in past, it was created from a grandparent cell with a birth-size $s_2$, a division-added-size $\Delta_{d_2}$, and a partitioning fraction $\beta_2$. Thus $s_1 = \beta_2(s_{2} + \Delta_{d_2})$, and $s_b = \beta_1(\Delta_{d_1} +\beta_2(\Delta_{d_2}+s_2))$. Moreover, the grandparent cell can be further tracked in the past, and this process goes on, and one can write:
\begin{equation}
    s_b \;=\;  \lim_{ n\rightarrow\infty}  \left[ \; \beta_1 \Delta_{d_1} \;+\; \beta_1 \beta_2 \Delta_{d_2} \;+\; \beta_1 \beta_2 \beta_3 \Delta_{d_3} \;+\; ... \;+\; \beta_1 \beta_2 ...\beta_n \Delta_{d_n} + \beta_1 \beta_2 ...\beta_n s_n \; \right].
\end{equation}
And more concisely, it can be written as:
\begin{equation}
    s_b = \lim_{ n\rightarrow\infty} \sum_{i=1}^{n} \left( \Delta_{d_i} \prod_{j=1}^{i} \beta_j \right) \;+\; \lim_{ n\rightarrow\infty} s_n\prod_{i=1}^{n} \beta_i.
\end{equation}
The second term can be neglected in the large $n$ limit because $\beta_i < 1$ and their product will result in further smaller quantity. Hence $s_b$ can be written as:
\begin{equation} \label{eq: sbRaw}
    s_b = \lim_{ n\rightarrow\infty} \sum_{i=1}^{n} \left( \Delta_{d_i} \prod_{j=1}^{i} \beta_j \right).
\end{equation}
This series can be generated from a recursion relation in $s_b$ that is given below:
\begin{equation}
    s_b' \;=\; \beta (\Delta_d + s_b),
\end{equation}
where one starts from a given $s_b$, and generates $s_b^{'}$ by sampling $\Delta_d$ and $\beta$ from their corresponding distributions $\Omega(\Delta_d)$ and $\kappa(\beta)$, respectively. In the next step of the recursion, $s_b^{'}$ becomes $s_b$, and new values of $\Delta_d$ and $\beta$ are sampled to obtain new $s_b^{'}$. This process goes on, and the series above can be obtained. One can further write from the recursion relation:
\begin{equation}
    s_b^{'^{n}} \;=\; \beta^n (\Delta_d + s_b)^n.
\end{equation}
Now multiplying the both sides by $\zeta(s_b^{'}) ds_b^{'}$ and integrating, we get:
\begin{equation}
    \int_0^{\infty} s_b^{'^{n}} \;\zeta(s_b^{'}) ds_b^{'} \;=\; \int_0^{\infty} \beta^n (\Delta_d + s_b)^n \;\zeta(s_b^{'}) ds_b^{'}.
\end{equation}
In the steady state, the distributions for both $s_b'$ and $s_b$ are the same, and are given by $\zeta(s_b)$. Hence,
\begin{equation}
    \int_0^{\infty} s_b^{'^{n}} \;\zeta(s_b^{'}) ds_b^{'} \;=\; \int_0^{\infty} \int_0^{\infty} \int_0^{1} \beta^n (\Delta_d + s_b)^n \;\zeta(s_b) \Omega(\Delta_d) \kappa(\beta) \;ds_b \;d\Delta_d \; d\beta.
\end{equation}
While writing the joint probability distribution for $\Delta_d$, $\beta$, and $s_b$ above, it is obvious that $\beta$ is taken to be independent of $\Delta_d$ and $s_b$, which is in line with our assumption regarding the independence of $\beta$ with other single-cell quantities. It can also be seen that $s_b$ is taken to be independent of $\Delta_d$, which is true only for the Adder model. Therefore, after expanding $(\Delta_d + s_b)^{n}$ into a binomial series, we can perform the integration over the three variables separately to get:
\begin{equation}
    \langle s_b^{'^n} \rangle \;=\; \langle \beta^n \rangle \sum_{i=0}^n \binom{n}{i} \langle s_b^i \rangle \langle \Delta_d^{n-i} \rangle.
\end{equation}
Again, in steady-state condition, the distributions for $s_b$ and $s_b^{'}$ are the same, therefore $\langle s_b^{'^n} \rangle  = \langle s_b^n \rangle$. Hence, from the equation above, one can obtain a recursion relation for $\langle s_b^{n} \rangle$:
\begin{equation} \label{eq: recursiveSbAdder}
    \langle s_b^n \rangle \;=\; \frac{\langle \beta^n \rangle \sum_{i=0}^{n-1} \binom{n}{i} \langle s_b^i \rangle \langle \Delta_d^{n-i} \rangle}{1-\langle \beta^n \rangle}.
\end{equation}
If first $n$ moments are known for $\Delta_d$ and $\beta$, one can also obtain first $n$ moments for $s_b$ using the recursion relation above. Note that, this recursion relation is true only for the Adder model because of the assumption regarding the independence of $s_b$ and $\Delta_d$. For the Adder model, $\kappa(\beta)$ and $\Omega(\Delta_d)$ are free-parameter distributions, and they are assumed to be known beforehand. This implies that all of the moments for $\Delta_d$ and $\beta$ are known, and hence one can calculate all the moments for $s_b$ from the recursion relation, in principle. But in practice, one calculates these moments for $s_b$ up to a finite order, and the full analytical distribution $\zeta(s_b)$ can be constructed using the method of characteristic function or maximum entropy distribution. However, the experimental distributions for $\Omega(\Delta_d)$ and $\kappa(\beta)$ are too noisy to allow reliable estimation of moments beyond the fourth order. Therefore, we calculated only first four moments for $s_b$ from the experimental data for $\beta$ and $\Delta_d$. After that, we found a Johnson SU or Johnson SB distribution with the values of its first four moments being very close to the calculated values. This was done using the moments matching algorithms for Johnson SU and SB distributions. The analytical form for $\zeta(s_b)$ was then taken to be that Johnson SU or Johnson SB distribution.

The relationship between the moments of $\Delta_d$ and $s_b$ is cumbersome for the Adder model. However, for the case of symmetric, non-stochastic partitioning ($\langle \beta \rangle = 0.5$ and $\sigma_{\beta}=0$) , the first four moments, namely mean, standard deviation, skewness, and excess kurtosis for $s_b$ can simply be calculated (from the recursion relation), and are given as:
\begin{equation}
    \langle s_b \rangle \;=\; \langle \Delta_d \rangle
\end{equation}
\begin{equation}
    \sigma_{s_b}^2 \;=\; \frac{\sigma_{\Delta_d}^2}{3}
\end{equation}
\begin{equation}
    \gamma_{s_b} = \frac{3\sqrt{3}}{7} \gamma_{\Delta_d}
\end{equation}
\begin{equation}
    K_{s_b} = \frac{3}{5} K_{\Delta_d}
\end{equation}
where $\sigma_{x}$, $\gamma_{x}$, and $K_{x}$ represent standard deviation, skewness, and excess kurtosis for a variable $x$.

\section{Obtaining remaining single-cell-level distributions} \label{sec: Sdistributions}

\subsection{Probability transformation of random variables} \label{subsec: SdistributionsProbTrans}
Suppose we are given a joint probability distribution of two random variables $x$ and $y$ as $P(x,y)$. The probability distribution for another variable $z=Z(x,y)$ is given as \cite{rileyHobsonBence} :
\begin{equation} \label{eq: probTransRaw}
    Q(z) \;=\; \int_{0}^{\infty} \int_{0}^{\infty} P(x,y) \delta(z-Z(x,y)) dx\;dy.
\end{equation}
Note that the lower limit of integration is $0$, not $-\infty$, because the variables we have considered are related to the age and size of the cells, and they can not have negative values. Additionally, for our purpose of probability transformations, the relationships between these variables look like: $\Delta_d = \alpha \tau_d$ and $\Delta_d = s_d - s_b$. As one can see, if one of the three variables is fixed, the mapping between the remaining two variables is is one-to-one and differentiable. Therefore, under these assumptions, $x$ can be written as a function of $y$ and $z$ as $x=X(y,z)$,  and the same goes for $y$ as $y=Y(x,z)$. Also, $Z(X(y,z),y) = 1$. Now using the property of Dirac delta function:
\begin{equation}
    \delta(f(x)) \;=\; \sum_{x_i} \frac{\delta(x-x_i)}{|f'(x_i)|} \;\;\; \forall f(x_i) = 0,
\end{equation}
where $x_i$ are the zeros of the function $f(x)$. We take $f(x) = z-Z(x,y)$, where $y$ and $z$ are treated as constants. There is only one zero of $f(x)$ given as $x = X(y,z)$ because $z = Z(X(y,z),y)$. Hence, one can see that:
\begin{equation}
    \delta(z-Z(x,y)) \;=\; \frac{\delta(x-X(y,z))}{\left |\frac{\partial Z(x,y)}{\partial x} \right |_{x=X(y,z)}}.
\end{equation}
Additionally, for such functions, we have:
\begin{equation}
    \left( \frac{\partial Z(x,y)}{\partial x} \right)^{-1} \;=\; \left( \frac{\partial X(y,z)}{\partial z} \right).
\end{equation}
Therefore, 
\begin{equation}
    \delta(z-Z(x,y)) \;=\; \delta(x-X(y,z))\left |\frac{\partial X(y,z)}{\partial z} \right |.
\end{equation}
Hence, Eq. \ref{eq: probTransRaw} can be finally modified to a much simpler form:
\begin{equation}
    Q(z) \;=\; \int_{0}^{\infty} \int_{0}^{\infty} P(x,y) \delta(x-X(y,z))\left |\frac{\partial X(y,z)}{\partial z} \right | dx\;dy.
\end{equation}
After integrating over $x$, one finally obtains:
\begin{equation} 
    Q(z) \;=\; \int_{0}^{\infty} P(X(z,y),y) \;\; \left |\frac{\partial X(y,z)}{\partial z} \right | \;\; dy .
\end{equation}
The same thing can be done for $y$ also and one can show that the distribution for $z$ is given by:
\begin{equation} \label{eq: probTrans}
\begin{split}
    Q(z) &\;=\; \int_{0}^{\infty} P(X(z,y),y) \;\; \left |\frac{\partial X(y,z)}{\partial z} \right | \;\; dy \\
    &\;=\; \int_{0}^{\infty} P(x,Y(x,z)) \;\; \left |\frac{\partial Y(z,x)}{\partial z} \right | \;\; dx.
\end{split}
\end{equation}
This equation is used to derive the probability distribution of a variable given as a function of two other variables whose joint probability distribution is known. Using similar arguments, one can further show that if the joint probability distribution $P(x,y,w)$ of three variables, $x$, $y$, and $w$ is known, the probability distribution of another variable $z=Z(x,y,w)$ (where the function $Z(x,y,w)$ is invertible in x, y, and w, and has only one zero for each variable) can also be found using the equation below:
\begin{equation} \label{eq: probTransMultiplevariables}
    \begin{split}
    Q(z) &\;=\; \int_{0}^{\infty} \int_{0}^{\infty} dx \; dy \; P(x,y,W(x,y,z)) \;\; \left |\frac{\partial W(x,y,z)}{\partial z} \right | \\
    &\;=\;  \int_{0}^{\infty} \int_{0}^{\infty} dy \; dw \; P(X(z,y,w),y,w) \;\; \left |\frac{\partial X(y,z,w)}{\partial z} \right | \\
    &\;=\;  \int_{0}^{\infty} \int_{0}^{\infty} dw \; dx \; P(x,Y(x,w,z),w) \;\; \left |\frac{\partial Y(x,w,z)}{\partial z} \right | .
    \end{split}
\end{equation}

\subsection{Sizer model} \label{subsec: SdistributionsSizer}
Apart from the free-parameter distributions, $\zeta(s_b)$ for the Sizer model is obtained from $\Xi(s_d)$ and $\kappa(\beta)$ using Eq. \ref{eq: zetaSizer}, as discussed in  Appendix \ref{sec: SbirthSize}. The other single-cell-level distributions for the Sizer model, i.e., $\Gamma(\tau_d)$ and $\Omega(\Delta_d)$, can also be obtained from $\Xi(s_d)$ and $\zeta(s_b)$ using probability transformations. We already have $\Xi(s_d)$ and $\zeta(s_b)$ for the Sizer model. From the relation $\Delta_d \;=\; s_d - s_b$, using probability transformation, we have:

\begin{equation} \label{eq: deltaDsizer}
\begin{split}
    \Omega(\Delta_d) \;&=\; \int_{0}^{\infty} d s_b \; \zeta(s_b) \; \Xi(\Delta_d+s_b) \\
    &=\; \int_{0}^{\infty} d s_d \; \Xi(s_d) \; \zeta(s_d-\Delta_d)
\end{split}
\end{equation}
The distribution $\Omega(\Delta_d)$ obtained from the equation above is valid for any growth type. It can be linear, exponential or any other type of growth. This is because $\Omega(\Delta_d)$ is obtained only from the relations which do not depend on the type of growth. In order to obtain the distribution $\Gamma(\tau_d)$, one has to know the type of growth. For exponential single-cell growth, we have the relation $\tau_d \;=\;  \ln{(s_d/s_b})/\alpha$. Therefore, using the probability transformations for three variables, one can write:
\begin{equation} \label{eq: tau_dSizerStochAlpha}
\begin{split}
    \Gamma(\tau_d) \;&=\; \int_{0}^{\infty} d\alpha \; \Lambda_{E}(\alpha)\; \alpha \; e^{\alpha \tau_d} \int_{0}^{\infty} d s_b \; \zeta(s_b) \; s_b \; \Xi(s_b e^{\alpha \tau_d}) \\
    &=\; \int_{0}^{\infty} d \alpha \; \Lambda_{E}(\alpha) \; \alpha \;e^{-\alpha \tau_d} \int_{0}^{\infty} d s_d \; \Xi(s_d) \; s_d \; \zeta(s_d e^{-\alpha \tau_d})
\end{split}
\end{equation}
where $\Lambda_{E}(\alpha)$ is the probability distribution for exponential growth rate. Similarly, for the case of linear growth rate, we can use the relation $\tau_d = (s_d - s_b)/\alpha$. Therefore, again using the probability transformations for three variables, one can write:
\begin{equation}  \label{eq: tauDsizer}
    \begin{split}
    \Gamma(\tau_d) \;=&\;\int_{0}^{\infty} d\alpha \; \Lambda_{L}(\alpha) \; \alpha \int_{0}^{\infty} ds_b \; \zeta(s_b) \; \Xi(s_b + \alpha \tau_d) \\
    &=\; \int_{0}^{\infty} d\alpha \; \Lambda_{L}(\alpha) \; \alpha \int_{0}^{\infty} ds_d \; \Xi(s_d) \; \zeta(s_d - \alpha \tau_d)
\end{split}
\end{equation}
where $\Lambda_{L}(\alpha)$ is the probability distribution for linear growth rate. Note that in the equations derived above, the joint probability distribution for $s_b$ and $s_d$ is taken simply as the multiplication of their individual probability distributions, which is valid here because both of these variables are independent of each other because of the assumptions of the Sizer model. Similarly, for the joint probability distribution of $s_b$, $s_d$, and $\alpha$ also, their individual probability distributions are multiplied. This reflects the fact that the growth rate is also treated as an independent quantity, which is also discussed in Section \ref{subsec: autocorrelations}.

\subsection{Adder model} \label{subsec: SdistributionsAdder}
Apart from the free-parameter distributions, $\zeta(s_b)$ for the Sizer model is obtained from $\Omega(\Delta_d)$ and $\kappa(\beta)$ using the arguments presented in Appendix \ref{sec: SbirthSize}. The other single-cell-level distributions for the Sizer model, i.e., $\Gamma(\tau_d)$ and $\Omega(\Delta_d)$, can also be obtained from $\Omega(\Delta_d)$ and $\zeta(s_b)$ using probability transformations. We already have $\Omega(\Delta_d)$ and $\zeta(s_b)$ for the Adder model. From the relation $s_d \;=\; \Delta_d + s_b$, using probability transformation, we have:
\begin{equation} \label{eq: sdAdder}
\begin{split}
\Xi(s_d) \;&=\; \int_{0}^{\infty} d\Delta_d \; \Omega(\Delta_d) \; \zeta(s_d -\Delta_d) \\
&=\; \int_{0}^{\infty} ds_b \; \zeta(s_b) \; \Omega(s_d -s_b)
\end{split}
\end{equation}
The distribution $\Xi(s_d)$ obtained from the equation above is valid for any growth type. It can be linear, exponential or any other type of growth. As explained earlier, it is because $\Xi(s_d)$ is obtained only from the relations which do not depend on the type of growth. In order to obtain the distribution $\Gamma(\tau_d)$, one has to know the type of growth. For exponential single-cell growth, we have the relation $\tau_d \;=\;  \ln{(1+(\Delta_d/s_b)})/\alpha$. Therefore, using the probability transformations for three variables, one can write:
\begin{equation} \label{eq: tauDadder}
\begin{split}
    \Gamma(\tau_d)\;&=\; \int_{0}^{\infty} d\alpha \; \Lambda_{E}(\alpha) \; \frac{\alpha e^{\alpha \tau_d}}{(e^{\alpha \tau_d}-1)^2} \int_{0}^{\infty} d\Delta_d \;\Delta_d\; \Omega(\Delta_d) \; \zeta \left(\frac{\Delta_d}{e^{\alpha \tau_d}-1} \right)  \\
    &=\; \int_{0}^{\infty} d\alpha \; \Lambda_{E}(\alpha) \; \alpha \;e^{\alpha \tau_d} \int_{0}^{\infty} ds_b \; s_b\; \zeta(s_b) \; \Omega(s_b(e^{\alpha \tau_d}-1))
\end{split}
\end{equation}
where $\Lambda_{E}(\alpha)$ is the probability distribution for exponential growth rate. Similarly, for the case of linear growth rate, we can use the relation $\tau_d = \Delta_d/\alpha$. Therefore, again using the probability transformations for three variables, one can write:
\begin{equation} \label{eq: tauDadderProbTrans}
\begin{split}
    \Gamma(\tau_d) \;&=\; \int_{0}^{\infty} d\alpha \; \Lambda_{L}(\alpha) \; \alpha \; \Omega(\alpha \tau_d) \\
    &=\; \int_{0}^{\infty} d\Delta_d \; \Omega(\Delta_d) \; \Lambda_{L}(\Delta_d/\tau_d) \; \frac{\Delta_d}{\tau_d^2} 
\end{split}
\end{equation}
where $\Lambda_{L}(\alpha)$ is the probability distribution for linear growth rate. Note that in the equations derived above, the joint probability distribution for $s_b$ and $\Delta_d$ is taken simply as the multiplication of their individual probability distributions, which is valid here because both of these variables are independent of each other because of the assumptions of the Adder model. Similarly, for the joint probability distribution of $s_b$, $\Delta_d$, and $\alpha$ also, their individual probability distributions are multiplied. This reflects the fact that the growth rate is also treated as an independent quantity, which is mentioned in Section \ref{subsec: autocorrelations}.

\section{Statistical relationships for single-cell quantities} \label{sec: Srelationships}

\subsection{Size-related quantities} \label{subsec: SrelationshipsSizeRelated}
First of all, let us derive the relationships between the mean of the size-related single-cell quantities. We have $s_b = \beta s_d$, where $s_b$ is the birth-size of the daughter cell and $s_d$ is the division-size of the parent cell. Taking average on both sides of this equation:
\begin{equation}
    \langle s_b \rangle = \langle \beta s_d \rangle
\end{equation}
Since the partitioning fraction $\beta$ is not correlated with other quantities, such as $s_d$ and $\Delta_d$ (as discussed in Section \ref{subsec: autocorrelations}), one can write:
\begin{equation} \label{eq: meanSbSd}
    \langle s_b \rangle = \langle \beta \rangle \langle s_d \rangle
\end{equation}
Further, from the equation, $\Delta_d = s_d - s_b$, we have $\langle \Delta_d \rangle = \langle s_d \rangle - \langle s_b \rangle$. Using the equation above, one can write:
\begin{equation} \label{eq: meanDeltaDsd}
    \langle \Delta_d \rangle = (1-\langle \beta \rangle) \langle s_d \rangle
\end{equation}
Eq. \ref{eq: meanSbSd} and \ref{eq: meanDeltaDsd} can be combined into one equation, i.e. Eq. \ref{eq: meanSbSdDeltaD}. Now let us derive the relationship between the standard deviation of $s_b$ and $s_d$. Using the equation $s_b = \beta s_d$ and Eq. \ref{eq: meanSbSd}, one can write:
\begin{equation}
    \langle (s_b - \langle s_b \rangle)^2 \rangle \;=\; \langle (\beta s_d - \langle \beta \rangle \langle s_d \rangle)^2 \rangle
\end{equation}
which can be simplified to obtain:
\begin{equation} \label{eq: sigmaSbSdRaw}
    \sigma_{s_b}^2 \;=\; \langle \beta \rangle ^2 \sigma_{s_d}^2 + \langle s_d \rangle^2 \sigma_{\beta}^2 + \sigma_{\beta}^2 \sigma_{s_d}^2
\end{equation}
Further modifying this equation easily gives us Eq. \ref{eq: sigmaSbSigmaSd}. Again, the relationships derived so far are valid for all growth and division models. Now, using the equation $\Delta_d = s_d - s_b$, and Eq. \ref{eq: meanDeltaDsd}, one can write:
\begin{equation}
    \langle (\Delta_d - \langle \Delta_d \rangle)^2 \rangle \;=\; \langle ((s_d - \langle s_d \rangle) - (s_b - \langle s_b \rangle))^2 \rangle
\end{equation}
which can be simplified as:
\begin{equation}
    \sigma_{\Delta_d}^2 \;=\; \sigma_{s_d}^2 + \sigma_{s_b}^2 - 2C(s_d,s_b)
\end{equation}
where $C(s_d,s_b) = \langle s_d s_b \rangle - \langle s_d \rangle \langle s_b \rangle$ is the correlation between $s_d$ and $s_b$. For the Sizer model, it is equal to zero. Whereas, for the Adder model, let us calculate this correlation:
\begin{equation} \label{eq: corrSdSbAdder}
\begin{split}
    C(s_d,s_b) \;&=\;  \langle s_d s_b \rangle - \langle s_d \rangle \langle s_b \rangle \\
    &=\; \langle (\Delta_d + s_b) s_b \rangle - \langle s_d \rangle \langle s_b \rangle \\
    &=\;  \langle \Delta_d s_b \rangle + \langle s_b^2 \rangle - \frac{\langle s_b \rangle^2}{\beta} \\
    &=\; \frac{1-\langle \beta \rangle}{\langle \beta \rangle} \langle s_b \rangle^2 + \langle s_b^2 \rangle - \frac{\langle s_b \rangle^2}{\beta} \\
    &=\; \langle s_b^2 \rangle - \langle s_b \rangle^2\\
    &=\; \sigma_{s_b}^2
\end{split}
\end{equation}
Note that in the calculation above, we have made use of the equation \ref{eq: meanSbSdDeltaD}. Additionally, we have also used the fact that $\langle \Delta_d s_b \rangle = \langle \Delta_d \rangle \langle s_b \rangle$ for the Adder model. Now, for the Sizer model, we can write:
\begin{equation} \label{eq: sigmaDeltaDsdSizerRaw}
    \sigma_{\Delta_d}^2 \;=\; \sigma_{s_d}^2 + \sigma_{s_b}^2
\end{equation}
Further using Eq. \ref{eq: sigmaSbSdRaw}, and the equation above, one can easily obtain Eq. \ref{eq: sigmaDeltaDsdSizer}. Similarly, for the Adder model, using the correlation derived above, one can write:
\begin{equation} \label{eq: sigmaDeltaDsdAdderRaw}
    \sigma_{\Delta_d}^2 \;=\; \sigma_{s_d}^2 - \sigma_{s_b}^2
\end{equation}
Further using Eq. \ref{eq: sigmaSbSdRaw}, and the equation above, one can easily arrive at Eq. \ref{eq: sigmaDeltaDsdAdder}. While deriving all these relationships, we have not mentioned anything about the single-cell growth types. Therefore, the relationships derived so far are valid for any growth type. It might be linear, exponential, or other growth type.

For the case of fixed symmetric partitioning ($\langle \beta \rangle = 0.5$ and $\sigma_{\beta} = 0$), when the cell division is binary and symmetric along with no stochasticity in partitioning,  the relationships become simpler. The relationship between the means of the size-related quantities is given as (despite the division model):
\begin{equation}
    \langle s_d \rangle \;=\; 2\langle s_b \rangle \;=\; 2\langle \Delta_d \rangle
\end{equation}
Whereas for the Sizer model, their standard deviations are related as:
\begin{equation}
    \sigma_{s_d} \;=\; 2\sigma_{s_b} \;=\; 2\sigma_{\Delta_d}/\sqrt{5},
\end{equation}
and for the Adder model, one can write:
\begin{equation}
    \sigma_{s_d} \;=\; 2\sigma_{s_b} \;=\; 2\sigma_{\Delta_d}/\sqrt{3}.
\end{equation}

\subsection{Division-time for exponential single-cell growth} \label{subsec: SrelationshipsDivisionTimeExpo}
Division-time is a function of $s_d$, $s_b$, and $\alpha$. $\alpha$ here corresponds to the exponential growth rate. For the case of exponential growth, it can be written as:
\begin{equation}
    \tau_d \;=\; \frac{1}{\alpha}\ln{\left(\frac{s_d}{s_b}\right)} .
\end{equation}
Using Taylor series expansion in the vicinity of the point $(\langle s_d \rangle,\langle s_b \rangle, \langle \alpha \rangle)$, we can write:
\begin{equation} \label{eq: tauDrawTaylorExpo}
\begin{split}
    \tau_d \;=\;& \frac{1}{\langle \alpha \rangle} \ln{\left(\frac{\langle s_d \rangle}{\langle s_b \rangle} \right)} \;+\; \frac{1}{\langle \alpha \rangle \langle s_d \rangle} (s_d - \langle s_d \rangle) \;-\; \frac{1}{\langle \alpha \rangle \langle s_b \rangle}(s_b - \langle s_b \rangle) \;-\; \frac{\ln{(\langle s_d \rangle /\langle s_b \rangle)}}{\langle \alpha \rangle^2} (\alpha - \langle \alpha \rangle) \\
    &  +\; \frac{\ln{(\langle s_d \rangle /\langle s_b \rangle)}}{\langle \alpha \rangle} \sum_{m=2}^{\infty} \frac{(-1)^m (\alpha -\langle \alpha \rangle)^m)}{\langle \alpha \rangle^m} + \frac{1}{\langle \alpha \rangle} \sum_{n=2}^{\infty} \frac{(-1)^{n+1}}{n} \left\{  \frac{(s_d -\langle s_d \rangle)^n}{\langle s_d \rangle^n} - \frac{(s_b -\langle s_b\rangle)^n}{\langle s_b \rangle^n} \right\} \\
    & +\; \sum_{n=1}^{\infty} \sum_{m=1}^{\infty} \frac{(-1)^{m+n+1}}{n \langle \alpha \rangle} \left\{  \frac{(s_d -\langle s_d \rangle)^n}{\langle s_d \rangle^n} - \frac{(s_b -\langle s_b\rangle)^n}{\langle s_b \rangle^n} \right\} \left( \frac{(\alpha - \langle \alpha \rangle)^m}{\langle \alpha \rangle^m} \right).
\end{split}
\end{equation}
Taking average of the equation above, we have,
\begin{equation}
\begin{split}
   \langle  \tau_d  \rangle\;=\;& \frac{1}{\langle \alpha \rangle} \ln{\left(\frac{\langle s_d \rangle}{\langle s_b \rangle} \right)} \; +\; \frac{\ln{(\langle s_d \rangle /\langle s_b \rangle)}}{\langle \alpha \rangle} \sum_{m=2}^{\infty} \frac{(-1)^m \mu_{m}(\alpha)}{\langle \alpha \rangle^m} + \frac{1}{\langle \alpha \rangle} \sum_{n=2}^{\infty} \frac{(-1)^{n+1}}{n} \left\{  \frac{\mu_n(s_d)}{\langle s_d \rangle^n} - \frac{\mu_(s_b)}{\langle s_b \rangle^n} \right\} \\
    & +\; \sum_{n=1}^{\infty} \sum_{m=1}^{\infty} \frac{(-1)^{m+n+1}}{n \langle \alpha \rangle} \left\{  \frac{\mu_n(s_d)}{\langle s_d \rangle^n} - \frac{\mu_n(s_b)}{\langle s_b \rangle^n} \right\} \left( \frac{\mu_m(\alpha)}{\langle \alpha \rangle^m} \right),
\end{split}
\end{equation}
where $\mu_n(x) = \langle (x-\langle x \rangle)^n \rangle$ is $n$th raw moment for the variable $x$. Note that we have treated $\alpha$ independent of $s_d$ and $s_b$, which we had discussed in Section \ref{subsec: autocorrelations}. Also note that $\mu_n(x)/\langle x \rangle = 0$ for $n=1$ because because $\langle x - \langle x \rangle \rangle /\langle x \rangle= 0$. We further assume that the distribution for $\alpha$, $s_d$, and $s_b$ are sharply spiked. This means that the higher order moments are negligible in comparison to the mean of the distribution. More precisely, the terms containing $\mu_n(x)/\langle x \rangle^n$ in the equations above vanish for $s_d$, $s_b$, and $\alpha$, for $n>1$ also. Hence, we obtain the following relation:
\begin{equation}
    \langle \tau_d \rangle \;=\; \frac{\ln(\langle s_d \rangle / \langle s_b \rangle)}{\langle \alpha \rangle} \;=\; -\frac{\ln(\langle \beta \rangle)}{\langle \alpha \rangle}
\end{equation}
which is just Eq. \ref{eq: meanTauDexponential}. Additionally, if one looks in the vicinity of the point $(\langle s_d \rangle,\langle s_b \rangle, \langle \alpha \rangle)$ such that $s_d - \langle s_d \rangle \ll \langle s_d \rangle$, $s_b - \langle s_b \rangle \ll \langle s_b \rangle$, and $\alpha - \langle \alpha \rangle \ll \langle \alpha \rangle$, considering only the first-order terms is sufficient (because the distributions for $\alpha$, $s_d$, and $\beta$ are sharply spiked). Hence Eq. \ref{eq: tauDrawTaylorExpo} becomes: 
\begin{equation} \label{eq: tauDtaylorExpoFirstOrder}
    \tau_d - \langle \tau_d \rangle = \frac{1}{\langle \alpha \rangle \langle s_d \rangle} (s_d - \langle s_d \rangle) \;-\; \frac{1}{\langle \alpha \rangle \langle s_b \rangle}(s_b - \langle s_b \rangle) \;-\; \frac{\langle \tau_d \rangle}{\langle \alpha \rangle} (\alpha - \langle \alpha \rangle)
\end{equation}
And therefore,
\begin{equation}
\begin{split}
    \langle (\tau_d - \langle \tau_d \rangle)^2 \rangle \;=\; \sigma_{\tau_d}^2 \;&=\; \frac{1}{\langle \alpha \rangle^2 \langle s_d \rangle^2} \sigma_{s_d}^2 \;+\; \frac{1}{\langle \alpha \rangle^2 \langle s_b \rangle^2}\sigma_{s_b}^2 \;+\; \frac{\langle \tau_d \rangle^2}{\langle \alpha \rangle^2} \sigma_{\alpha}^2  \\ 
    & - \frac{2}{\langle \alpha \rangle^2 \langle s_d \rangle \langle s_b \rangle} \langle (s_d -\langle s_d\rangle)(s_b - \langle s_b \rangle) \rangle - \frac{2 \langle \tau_d \rangle}{\langle \alpha \rangle^2 \langle s_d \rangle} \langle (\alpha -\langle \alpha \rangle)(s_d - \langle s_d \rangle) \rangle  \\
    & + \frac{2 \langle \tau_d \rangle}{\langle \alpha \rangle^2 \langle s_b \rangle} \langle (\alpha -\langle \alpha \rangle)(s_b - \langle s_b \rangle) \rangle
\end{split}
\end{equation}
Since there is no correlation between $\alpha$ and $s_d$, and $\alpha$ and $s_b$, the cross terms containing $\langle (\alpha -\langle \alpha \rangle)(s_d - \langle s_d \rangle) \rangle $ and $\langle (\alpha -\langle \alpha \rangle)(s_b - \langle s_b \rangle) \rangle $ will vanish. Therefore, the equation is simplified as:
\begin{equation} \label{eq: COVtauStochExpoRaw}
    \langle (\tau_d - \langle \tau_d \rangle)^2 \rangle \;=\; \sigma_{\tau_d}^2 \;=\; \frac{1}{\langle \alpha \rangle^2 \langle s_d \rangle^2} \sigma_{s_d}^2 \;+\; \frac{1}{\langle \alpha \rangle^2 \langle s_b \rangle^2}\sigma_{s_b}^2 \;+\; \frac{\langle \tau_d \rangle^2}{\langle \alpha \rangle^2} \sigma_{\alpha}^2  - \frac{2C(s_d,s_b)}{\langle \alpha \rangle^2 \langle s_d \rangle \langle s_b \rangle}
\end{equation}
For the Sizer model, $C(s_d,s_b)=0$, therefore one can write:
\begin{equation}
    \sigma_{\tau_d}^2 \;=\; \frac{1}{\langle \alpha \rangle^2 \langle s_d \rangle^2} \sigma_{s_d}^2 \;+\; \frac{1}{\langle \alpha \rangle^2 \langle s_b \rangle^2}\sigma_{s_b}^2 \;+\; \frac{\langle \tau_d \rangle^2}{\langle \alpha \rangle^2} \sigma_{\alpha}^2
\end{equation}
Dividing the equation above by $\langle \tau_d \rangle^2$ and using Eq. \ref{eq: meanTauDexponential}, one can obtain Eq. \ref{eq: COVtauDexpSizer}. Additionally, if we are talking about the Adder model, as derived earlier $C(s_d,s_b) = \sigma_{s_b}^2$. Hence, for the Adder model, one can write:
\begin{equation}
    \sigma_{\tau_d}^2 \;=\; \frac{1}{\langle \alpha \rangle^2 \langle s_d \rangle^2} \sigma_{s_d}^2 \;+\; \frac{(1-2\langle \beta \rangle)}{\langle \alpha \rangle^2 \langle s_b \rangle^2}\sigma_{s_b}^2 \;+\; \frac{\langle \tau_d \rangle^2}{\langle \alpha \rangle^2} \sigma_{\alpha}^2
\end{equation}
Similarly, dividing the equation above by $\langle \tau_d \rangle^2$ and using Eq. \ref{eq: meanTauDexponential}, one can obtain Eq. \ref{eq: COVtauDexpAdder}.

As mentioned earlier, here too, the relationships become simpler for the case of fixed symmetric partitioning. For mean division-time, one can write (despite the division model):
\begin{equation}
    \langle \tau_d \rangle \;=\; \frac{\ln{(2)}}{\langle \alpha \rangle}.
\end{equation}
Whereas for the Sizer model, one obtains the following expression for $\sigma_{\tau_d}$:
\begin{equation}
\begin{split}
     \frac{\sigma_{\tau_d}}{\langle \tau_d \rangle} \;=\; \sqrt{\left( \frac{\sqrt{2} \sigma_{s_b}}{\ln{(2)} \langle s_b \rangle} \right)^2 + \left( \frac{\sigma_{\alpha}}{\langle \alpha \rangle} \right)^2} \;=\; \sqrt{\left( \frac{\sqrt{2} \sigma_{s_d}}{\ln{(2)} \langle s_d \rangle} \right)^2 + \left( \frac{\sigma_{\alpha}}{\langle \alpha \rangle} \right)^2} \;=\; \sqrt{\left( \frac{\sqrt{2} \sigma_{\Delta_d}}{\sqrt{5} \ln{(2)} \langle \Delta_d \rangle} \right)^2 + \left( \frac{\sigma_{\alpha}}{\langle \alpha \rangle} \right)^2} 
\end{split}
\end{equation}
And for the Adder model, the expression is slightly changed as:
\begin{equation}
\begin{split}
     \frac{\sigma_{\tau_d}}{\langle \tau_d \rangle} \;=\; \sqrt{\left( \frac{ \sigma_{s_b}}{\ln{(2)} \langle s_b \rangle} \right)^2 + \left( \frac{\sigma_{\alpha}}{\langle \alpha \rangle} \right)^2} \;=\; \sqrt{\left( \frac{\sigma_{s_d}}{\ln{(2)} \langle s_d \rangle} \right)^2 + \left( \frac{\sigma_{\alpha}}{\langle \alpha \rangle} \right)^2} \;=\; \sqrt{\left( \frac{\sigma_{\Delta_d}}{\sqrt{3} \ln{(2)} \langle \Delta_d \rangle} \right)^2 + \left( \frac{\sigma_{\alpha}}{\langle \alpha \rangle} \right)^2} 
\end{split}
\end{equation}

\subsection{Division-time for linear single-cell growth} \label{subsec: SrelationshipsDivisionTimeLinear}
For linear growth, we have $\tau_d = \Delta_d /\alpha$, where $\alpha$ is the linear growth rate. Since, both $\Delta_d$ and $\alpha$ are stochastic quantities, we expand the function $\tau_d$ about the point$(\langle \Delta_d \rangle ,\langle \alpha \rangle)$ using Taylor series expansion in two variables:
\begin{equation}
    \tau_d \;=\; \frac{\langle \Delta_d \rangle}{\langle \alpha \rangle} \;+\; \sum_{n=1}^{\infty} (-1)^n \left( \frac{\langle \Delta_d \rangle}{\langle \alpha \rangle} \right) \frac{(\alpha - \langle \alpha \rangle)^n}{\langle \alpha \rangle^n}  \;+\; \sum_{n=0}^{\infty} (-1)^n \left( \frac{\langle \Delta_d \rangle}{\langle \alpha \rangle} \right)  \left( \frac{\Delta_d - \langle \Delta_d \rangle}{\langle \Delta_d \rangle} \right) \frac{(\alpha - \langle \alpha \rangle)^n}{\langle \alpha \rangle^n}.
\end{equation}
Taking average on both sides of the above equation, we get:
\begin{equation}
    \langle \tau_d \rangle \;=\; \frac{\langle \Delta_d \rangle}{\langle \alpha \rangle} \;+\; \sum_{n=1}^{\infty} (-1)^n \left( \frac{\langle \Delta_d \rangle}{\langle \alpha \rangle} \right) \frac{\mu_n(\alpha)}{\langle \alpha \rangle^n}  \;+\; \sum_{n=0}^{\infty} (-1)^n \left( \frac{\langle \Delta_d \rangle}{\langle \alpha \rangle} \right)  \left( \frac{\langle \Delta_d - \langle \Delta_d \rangle \rangle}{\langle \Delta_d \rangle} \right) \frac{\mu_n(\alpha)}{\langle \alpha \rangle^n},
\end{equation}
where we have used the fact that $\Delta_d$ and $\alpha$ are independent of each other, which we have discussed in Section \ref{subsec: autocorrelations}. The term corresponding to the second summation evaluates to zero because $\langle \Delta_d - \langle \Delta_d \rangle \rangle = 0$. Further, if we assume that the distribution of $\alpha$ is sharply spiked, i.e.  $\mu_n(\alpha)/\langle \alpha \rangle^n \approx 0$ for $n \geq 2$, the term corresponding to the first summation also evaluates to zero. Note that $\mu_n(\alpha)/\langle \alpha \rangle = 0$ for $n=1$ already because $\langle \alpha - \langle \alpha \rangle \rangle = 0$. Therefore, we obtain:
\begin{equation}
    \langle \tau_d \rangle \;=\; \frac{\langle \Delta_d \rangle}{\langle \alpha \rangle}
\end{equation}
Additionally, while we look in the close neighborhood of the point $(\langle \Delta_d \rangle ,\langle \alpha \rangle)$ such that $\Delta_d - \langle \Delta_d \rangle \ll \langle \Delta_d \rangle $ and $\alpha - \langle \alpha \rangle \ll \langle \alpha \rangle $, higher order terms, being very small, do not contribute much. Therefore, we consider the first-order terms only and write for $\tau_d$ as:
\begin{equation}
    \tau_d - \langle \tau_d \rangle =  \frac{1}{\langle \alpha \rangle}(\Delta_d - \langle \Delta_d \rangle) - \frac{\langle \Delta_d \rangle}{\langle \alpha \rangle^2}(\alpha -\langle \alpha \rangle) 
\end{equation}
Therefore,
\begin{equation}
    \langle (\tau_d - \langle \tau_d \rangle)^2 \rangle =  \frac{1}{\langle \alpha \rangle^2} \langle (\Delta_d - \langle \Delta_d \rangle)^2 \rangle + \frac{\langle \Delta_d \rangle^2}{\langle \alpha \rangle^4} \langle (\alpha -\langle \alpha \rangle)^2 \rangle - \frac{2 \langle \Delta_d \rangle}{\langle \alpha \rangle^3} \langle (\Delta_d - \langle \Delta_d \rangle)(\alpha - \langle \alpha \rangle) \rangle
\end{equation}
Since $\Delta_d$ and $\alpha$ are not correlated. Therefore, the third term on the RHS of the equation above vanishes. Further, dividing this equation by $\langle \tau_d \rangle^2$, and using Eq. \ref{eq: meanTauDlinear}, one can write:
\begin{equation}
    \frac{\sigma_{\tau_d}}{\langle \tau_d \rangle} \;=\; \sqrt{\left( \frac{\sigma_{\Delta_d}}{\langle \Delta_d \rangle} \right)^2 + \left( \frac{\sigma_{\alpha}}{\langle \alpha \rangle} \right)^2}
\end{equation}
which can simply be written as:
\begin{equation} \label{eq: COVtauDlinearRaw}
    C(\tau_d) \;=\; \sqrt{C(\Delta_d)^2 + C(\alpha)^2}
\end{equation}
This equation holds for both the Sizer and Adder model. However, while we write it in terms of $s_d$ and $s_b$ instead of $\Delta_d$, it will be different for the Sizer and Adder models. For the Sizer model, using Eq. \ref{eq: sigmaDeltaDsdSizerRaw} and \ref{eq: meanSbSdDeltaD}, one can write:
\begin{equation}
    C(\Delta_d)^2 \;=\; \frac{C(s_d)^2 + \langle \beta \rangle^2 C(s_b)^2 }{(1-\langle \beta \rangle)^2}
\end{equation}
Similarly, for the Adder model using Eq. \ref{eq: sigmaDeltaDsdAdderRaw} and \ref{eq: meanSbSdDeltaD}, one can write for $C(\Delta_d)$:
\begin{equation}
    C(\Delta_d)^2 \;=\; \frac{C(s_d)^2 - \langle \beta \rangle^2 C(s_b)^2 }{(1-\langle \beta \rangle)^2}
\end{equation}
Further putting the values of $C(\Delta_d)$ obtained for the Sizer and Adder models into Eq. \ref{eq: COVtauDlinearRaw}, one can obtain Eq. \ref{eq: COVtauDlinearSizer} and \ref{eq: COVtauDlinearAdder}.

As mentioned earlier, here too, the relationships and expressions become simpler for the case of fixed symmetric partitioning. For this case with linear single-cell growth, one obtains for mean division-time:
\begin{equation}
    \langle \tau_d \rangle \;=\; \frac{\langle \Delta_d \rangle}{\langle \alpha \rangle} \;=\; \frac{\langle s_b \rangle}{\langle \alpha \rangle} \;=\; \frac{\langle s_d \rangle}{2\langle \alpha \rangle}
\end{equation}
For the Sizer model, one obtains the following expression for $\sigma_{\tau_d}$:
\begin{equation}
    \frac{\sigma_{\tau_d}}{\langle \tau_d \rangle} \;=\; \sqrt{\left( \frac{\sigma_{\Delta_d}}{\langle \Delta_d \rangle} \right)^2 + \left( \frac{\sigma_{\alpha}}{\langle \alpha \rangle} \right)^2} \;=\; \sqrt{\left( \frac{\sqrt{5} \sigma_{s_d}}{\langle s_d \rangle} \right)^2 + \left( \frac{\sigma_{\alpha}}{\langle \alpha \rangle} \right)^2} \;=\; \sqrt{\left( \frac{\sqrt{5} \sigma_{s_b}}{\langle s_b \rangle} \right)^2 + \left( \frac{\sigma_{\alpha}}{\langle \alpha \rangle} \right)^2}
\end{equation}
Whereas, for the Adder model, this relationship is slightly changed as:
\begin{equation}
    \frac{\sigma_{\tau_d}}{\langle \tau_d \rangle} \;=\; \sqrt{\left( \frac{\sigma_{\Delta_d}}{\langle \Delta_d \rangle} \right)^2 + \left( \frac{\sigma_{\alpha}}{\langle \alpha \rangle} \right)^2} \;=\; \sqrt{\left( \frac{\sqrt{3} \sigma_{s_d}}{\langle s_d \rangle} \right)^2 + \left( \frac{\sigma_{\alpha}}{\langle \alpha \rangle} \right)^2} \;=\; \sqrt{\left( \frac{\sqrt{3} \sigma_{s_b}}{\langle s_b \rangle} \right)^2 + \left( \frac{\sigma_{\alpha}}{\langle \alpha \rangle} \right)^2}
\end{equation}

\section{Correlations}\label{sec: Scorrelations}

In the derivations below, $C(a,b) = \langle a b \rangle - \langle a \rangle \langle b\rangle$ represents the correlation between the quantities $a$ and $b$ and $P(a,b) = C(a,b)/(\sigma_a \sigma_b) = \langle a b \rangle - \langle a \rangle \langle b\rangle / (\sigma_a \sigma_b)$ represents the Pearson Correlation Coefficient between the quantities $a$ and $b$. We make extensive use of Eq. \ref{eq: meanSbSdDeltaD} in these derivations.

\subsection{Sizer}
For the Sizer model, division-size of a cell $(s_d)$ is selected independently of its birth-size $(s_b)$, and the division-sizes of cells from preceding generations in the lineage. Therefore, we have:
\begin{equation}
    C(s_{d_i} ,s_{d_j}) \;=\; \langle s_{d_i} s_{d_j} \rangle - \langle s_d \rangle ^2 = \delta_{ij} \sigma_{s_d}^2,
\end{equation}
where $s_{d_i}$, and $s_{d_j}$ are the division-sizes for the $i$th and $j$th generation cells. And,
\begin{equation}
    C(s_d , s_b) \;=\; \langle s_d s_b\rangle - \langle s_d \rangle \langle s_b \rangle \;=\; 0.
\end{equation}
Hence $P(s_d,s_b) = 0$ because $C(s_d,s_b) = 0$.  Now, the correlation between $\Delta_d$ and $s_b$ can be derived as:
\begin{equation}
\begin{split}
    C(\Delta_d,s_b) \;&=\; \langle \Delta_d s_b \rangle - \langle \Delta_d \rangle \langle s_b \rangle \\
    &=\; \langle (s_d-s_b) s_b \rangle - \langle \Delta_d \rangle \langle s_b \rangle \\
    &=\; \langle s_d s_b \rangle - \langle s_b^2 \rangle - \langle \Delta_d \rangle \langle s_b \rangle \\
    &=\; \langle s_d \rangle \langle s_b \rangle - \langle s_b^2 \rangle - \langle \Delta_d \rangle \langle s_b \rangle \\
     &=\; \frac{\langle s_b \rangle^2}{\langle \beta \rangle} - \langle s_b^2 \rangle - (\frac{\langle s_b \rangle}{\langle \beta \rangle} -\langle s_b \rangle )\langle s_b \rangle \\
     &=\; -\sigma_{s_b}^2
\end{split}
\end{equation}
Therefore,
\begin{equation}
\begin{split}
    P(\Delta_d,s_b) \;&=\; C(\Delta_d,s_b)/(\sigma_{s_d} \sigma_{s_b}) \\
    &=\; \frac{- \sigma_{s_b}^2}{\sigma_{s_b} \sigma_{\Delta_d}} \\
    &=\; \frac{-\sigma_{s_b}}{\sigma_{\Delta_d}}
\end{split}
\end{equation}
This expression can further be obtained only in terms of $\sigma_{s_d}$, $\langle s_d \rangle$, $\sigma_{\beta}$, and $\langle \beta \rangle$ (only in terms of the statistical features of free-parameter distributions) by writing both $\sigma_{s_b}$ and $\sigma_{\Delta_d}$ in terms of these quantities using Eq. \ref{eq: sigmaSbSigmaSd} and \ref{eq: sigmaDeltaDsdSizer}.

For the case of exponential growth, one can write: $C(\tau_d,s_b) = \langle \tau_d  s_b  \rangle - \langle \tau_d \rangle \langle s_b \rangle$. For $\tau_d$, one can use Eq. \ref{eq: tauDtaylorExpoFirstOrder} (with the assumption made in Section \ref{subsec: SrelationshipsDivisionTimeExpo}), and write for the correlation:
\begin{equation} \label{eq: corrTauDsbExpo}
\begin{split}
    C(\tau_d,s_b) \;&=\; \langle s_b \frac{1}{\alpha} \ln{(s_d/s_b)} \rangle - \langle s_b \rangle \langle \tau_d \rangle \\
    &=\; \langle s_b \left[ \langle \tau_d \rangle + \frac{s_d - \langle s_d \rangle}{\langle \alpha \rangle \langle s_d \rangle} - \frac{s_b - \langle s_b \rangle}{\langle \alpha \rangle \langle s_b \rangle} - \frac{\langle \tau_d \rangle} {\langle \alpha \rangle} (\alpha - \langle \alpha \rangle)\right] \rangle - \langle s_b \rangle \langle \tau_d \rangle\\
    &=\; \langle s_b \rangle \langle \tau_d \rangle + \langle \frac{s_b s_d - s_b \langle s_d \rangle}{\langle \alpha \rangle \langle s_d \rangle} \rangle - \langle \frac{s_b^2 - s_b \langle s_b \rangle}{\langle \alpha \rangle \langle s_b \rangle} \rangle - \frac{\langle \tau_d \rangle} {\langle \alpha \rangle} \langle (s_b \alpha - s_b \langle \alpha \rangle)  \rangle - \langle s_b \rangle \langle \tau_d \rangle\\
    &=\; \frac{C(s_d,s_b)}{\langle \alpha \rangle \langle s_d \rangle} - \frac{\sigma_{s_b}^2}{\langle \alpha \rangle \langle s_b \rangle}
\end{split} 
\end{equation}
For the Sizer model, $C(s_d,s_b) = 0$, therefore one obtains $C(\tau_d,s_b) = -\sigma_{s_b}^2/(\langle \alpha \rangle \langle s_b \rangle)$. Hence, $P(\tau_d,s_b) = -\sigma_{s_b}/(\langle \alpha \rangle \langle s_b \rangle \sigma_{\tau_d})$. Similarly, for the case of linear growth,
\begin{equation}
\begin{split}
    C(\tau_d,s_b) \;&=\; \langle \tau_d s_b \rangle - \langle \tau_d \rangle \langle s_b \rangle \\
    &=\; \langle \frac{(s_d-s_b)}{\alpha} s_b \rangle - \langle \frac{(s_d-s_b)}{\alpha} \rangle \langle s_b \rangle \\
    &=\; \langle \frac{s_d s_b}{\alpha} \rangle - \langle \frac{s_b^2}{\alpha} \rangle - \langle \frac{s_d}{\alpha} \rangle \langle s_b \rangle + \langle \frac{s_b}{\alpha} \rangle \langle s_b \rangle \\
\end{split}
\end{equation}
$\alpha$ is not correlated with $s_b$ and $s_d$. And for the Sizer model, $s_d$ is not correlated with $s_b$. Therefore, one can simply write:
\begin{equation}
\begin{split}
    C(\tau_d,s_b) \;&=\; \langle \frac{1}{\alpha} \rangle \left( \langle s_d \rangle \langle s_b \rangle - \langle s_b^2 \rangle - \langle s_d \rangle \langle s_b \rangle  + \langle s_b \rangle^2  \right) \\
    &=\; -\langle \frac{1}{\alpha} \rangle \sigma_{s_b}^2
\end{split}
\end{equation}
Now using Taylor series expansion, one can further prove that:
\begin{equation}
\begin{split}
    \langle \frac{1}{\alpha} \rangle \;&=\; \frac{1}{\langle \alpha \rangle} \sum_{n=0}^{\infty} \frac{(-1)^n \langle(\langle \alpha - \langle \alpha \rangle)^n \rangle}{\langle \alpha \rangle^n} \\
    &\approx \frac{1}{\langle \alpha \rangle} \left[ 1+ \frac{\sigma_{\alpha}^2}{\langle \alpha \rangle^2} + ... \right]
\end{split}
\end{equation}
Since the distribution for alpha is sharply spiked, the higher order terms will be negligible, therefore $C(\tau_d,s_b) = -\sigma_{s_b}^2/\langle \alpha \rangle$. Hence, $P(\tau_d,s_b) = -\sigma_{s_b}/( \sigma_{\tau_d}\langle \alpha \rangle)$. Again note that the using the relationships mentioned in Section \ref{subsec: relationships}, the final expressions for these correlations can be obtained only in terms of mean and standard deviations of the free-parameter distributions. Further, $\alpha$ in the expressions for the Pearson Correlation Coefficients between $s_b$ and $\tau_d$ for the case of exponential and linear growth is not the same quantity. For the first case, it is exponential growth rate, while for the other it is linear growth rate.

\subsection{Adder}
For the Adder model, division-added-size of a cell $(\Delta_d)$ is selected independently of its birth-size $(s_b)$, and the division-added-sizes of cells from preceding generations in the lineage. Therefore, we have:
\begin{equation}
    C(\Delta_{d_i} ,\Delta_{d_j}) \;=\; \langle \Delta_{d_i} \Delta_{d_j} \rangle - \langle \Delta_d \rangle ^2 = \delta_{ij} \sigma_{\Delta_d}^2,
\end{equation}
where $\Delta_{d_i}$, and $\Delta_{d_j}$ are the division-added-sizes for the $i$th and $j$th generation cells. And,
\begin{equation}
    C(\Delta_d , s_b) \;=\; \langle \Delta_d s_b\rangle - \langle \Delta_d \rangle \langle s_b \rangle \;=\; 0.
\end{equation}
Hence $P(\Delta_d,s_b) = 0$ because $C(\Delta_d,s_b) = 0$. Additionally, we have already derived the correlation between $s_d$ and $s_b$ (Eq. \ref{eq: corrSdSbAdder}) in Section \ref{subsec: SrelationshipsSizeRelated}. This is given as $C(s_d,s_b) = \sigma_{s_b}^2$ and thus Pearson Correlation Coefficient is given by $P(s_d,s_b) = C(s_d,s_b)/(\sigma_{s_d} \sigma_{s_b}) = \sigma_{s_b}/\sigma_{s_d}$. This expression can further be obtained only in terms of $\sigma_{\Delta_d}$, $\langle \Delta_d \rangle$, $\sigma_{\beta}$, and $\langle \beta \rangle$ (only in terms of the statistical features of free-parameter distributions) by writing both $\sigma_{s_b}$ and $\sigma_{s_d}$ in terms of these quantities using Eq. \ref{eq: meanSbSdDeltaD}, \ref{eq: sigmaSbSigmaSd}, and \ref{eq: sigmaDeltaDsdAdder}.

For the case of exponential growth, one can write: $C(\tau_d,s_b) = \langle \tau_d  s_b  \rangle - \langle \tau_d \rangle \langle s_b \rangle$. Further, one can use Eq. \ref{eq: corrTauDsbExpo} derived in the previous subsection in order to obtain this correlation for the Adder model as well. One can use the correlation $C(s_d,s_b) = \sigma_{s_b}^2$ and put it into Eq. \ref{eq: corrTauDsbExpo} to obtain $C(\tau_d,s_b) = \sigma_{s_b}^2 (\langle \beta \rangle -1)/(\langle \alpha \rangle \langle s_b \rangle)$. Hence, $P(\tau_d,s_b) = \sigma_{s_b} (\langle \beta \rangle -1)/(\langle \alpha \rangle \langle s_b \rangle \sigma_{\tau_d})$ for exponential growth. Similarly, for the case of linear growth:
\begin{equation}
\begin{split}
    C(\tau_d,s_b) \;&=\; \langle \tau_d s_b \rangle - \langle \tau_d \rangle \langle s_b \rangle \\
    &=\; \langle \frac{\Delta_d s_b}{\alpha}\rangle -  \langle \frac{\Delta_d}{\alpha} \rangle \langle s_b \rangle
\end{split}
\end{equation}
Now, $\alpha$ is not correlated with $\Delta_d$ and $s_b$. Additionally, for the Adder model, $\Delta_d$ is not correlated with $s_b$. Therefore, the expression above can be further simplified as:
\begin{equation}
\begin{split}
    C(\tau_d,s_b) \;&=\; \langle \frac{1}{\alpha} \rangle \left[ \langle \Delta_d s_b \rangle - \langle \Delta_d \rangle \langle s_b \rangle\right] \\
    &=\; \langle \frac{1}{\alpha} \rangle \left[ \langle \Delta_d \rangle \langle s_b \rangle - \langle \Delta_d \rangle \langle s_b \rangle\right] \\
    &=\; 0
\end{split}
\end{equation}
Hence, $P(\tau_d,s_b) = 0$ for linear growth. Again note that the using the relationships mentioned in Section \ref{subsec: relationships}, the final expressions for these correlations can be obtained in terms for the mean and standard deviations of the free-parameter distributions.

\section{The Timer model with linear growth type}\label{sec: StimerLinear}
For the Timer model of cell division, the division-time for the cells is constant. Additionally, if the single-cell growth is linear, one can write $s_d = s_b + \alpha \tau_d$, and thus $\Delta_d = \alpha \tau_d$. Therefore, a constant division-time implies a constant division-added-size, which is, in principal, the Adder model. Additionally, for the Timer model, $s_b$ is independent of $\tau_d$, such that $C(s_b, \tau_d) = \langle s_b \tau_d \rangle - \langle s_b \rangle \langle \tau_d \rangle = 0$. Further, the correlation between $s_b$ and $\Delta_d$ can be found as:
\begin{equation}
\begin{split}
    C(s_b,\Delta_d) \;&=\;  \langle s_b \Delta_d \rangle - \langle s_b\rangle \langle \Delta_d \rangle \\
    &=\; \langle s_b \alpha \tau_d \rangle - \langle s_b \rangle \langle \alpha \tau_d \rangle \\
\end{split}
\end{equation}
$\alpha$ is assumed to be independent of size-related quantities, such as $s_b$. Also, $\alpha$ and $\tau_d$ are independent for the Timer model ($\tau_d$ is the principal quantity now and is independent of the other free parameters). For the Timer model, $s_b$ and $\tau_d$ are also uncorrelated. Therefore, the correlation above evaluates to zero. This shows that the Timer and Adder models are equivalent for the case of linear growth. Similarly, the other correlations which are derived for the Adder model under linear growth are also the same for the Timer model. Furthermore, for the Timer model, the principal distribution is $\Gamma(\tau_d)$. One can find out $\Omega(\Delta_d)$ for the Timer model using the relationship $\Delta_d = \alpha \tau_d$ and probability transformation of random variables. This is given as:
\begin{equation}
\begin{split}
    \Omega(\Delta_d) \;&=\; \int_{0}^{\infty} d\alpha \; \frac{\Lambda_{L} (\alpha)}{\alpha} \; \Gamma \left(\frac{\Delta_d}{\alpha} \right)  \\
    &=\; \int_{0}^{\infty} d\tau_d \frac{\Gamma(\tau_d)}{\tau_d} \Lambda_{L} \left(\frac{\Delta_d}{\tau_d} \right)
\end{split}
\end{equation}
which is equivalent to Eq. \ref{eq: tauDadderProbTrans}. The other distributions, such as $\zeta(s_b)$, and $\Xi(s_d)$ can be further found by considering it to be an Adder model, for which these distributions have been obtained in Appendix \ref{sec: SbirthSize} and \ref{sec: Sdistributions}. 

\section{Extra results} \label{sec: SextraResults}

\subsection{Explanation for two separate histograms for the same experimental data} \label{subsec: SextraResultsSeparateHistograms}
In Fig. \ref{fig:tanouchiDisbns25C}, there are two separate histograms (for $\zeta(s_b)$, $\Xi(s_d)$, and $\Omega(\Delta_d)$) for the same experimental data. The only visible difference is the difference between single-cell growth types. One of them is for exponential growth, whereas the other is for linear growth. The same time-series data for the size of a single cell tracked over many generations gives different values of $s_b$, $s_d$, and $\Delta_d$ for each generation, when it is interpreted for different growth types.

For example, let us consider a sample cell-size data \cite{tanouchi2017} recorded at the intervals of one minute for a cell tracked over several generations in a mother-machine-type setup (the sizes are in $\mu$m):
\begin{equation}
    ..., 1.38127, 1.41607, 1.53428, ..., 3.539, 3.55245, 1.67728, 1.7095, ...
\end{equation}
From this sample data, one can easily see that the cell division has taken place between the events of recording the cell size to be $3.55245$ $\mu$m and $1.67728$ $\mu$m. Although one can consider $3.55245$ $\mu$m to be the division-size of the parent cell and $1.67728$ $\mu$m to be the birth-size of the daughter cell, the actual division-size is bigger than $3.55245$ $\mu$m, and the actual birth-size is smaller than $1.67728$ $\mu$m. This is because the division has taken place between the two events of recording these sizes. Therefore, the parent cell had extended itself a little more than $3.55245$ $\mu$m in that interval, and the daughter cell would have been born a little bit smaller and extended itself in that time interval for its size to be recorded as $1.67728$ $\mu$m.

In order to make these data points more precise, we assume that the division has taken place exactly in the middle of the these two events, and the cells grow exponentially, such that the division-size can be corrected as $s_d^{'} = s_d \exp{(\alpha_p t/2)}$, where $s_d$ is the division-size without any correction, $\alpha_p$ is the exponential growth rate of the parent cell, and $t$ is the least count for the time (the smallest time intervals after which the values of cell size are recorded). In the experiment performed by Tanouchi et. al. \cite{tanouchi2017}, the least count for the time is $1$ min. Similarly, the birth-size of the daughter cell can be corrected as: $s_b^{'} = s_b \exp{(-\alpha_d t/2)}$, where $\alpha_d$ is the exponential growth rate of the daughter cell. The values of the growth rates $\alpha_p$ and $\alpha_d$ can be calculated from the time-series data itself by fitting the data points linearly on a graph that has logarithm of cell size on y-axis and age of the cell on x-axis. Whereas, for the case of linear growth, these corrections are given as: $s_d^{'} = s_d + \alpha_p t/2$ and $s_b^{'} = s_b - \alpha_d t/2$, where $\alpha_p$ and $\alpha_d$ are linear growth rates for the parent and daughter cells. These growth rates can further be calculated from the time-series data by fitting the data points linearly on a graph that has cell size on y-axis and age of the cell on x-axis.

One can easily see that, while doing the corrections, $s_d$ and $s_b$ are extrapolated differently for exponential and linear growth types. Therefore, after the correction, $s_d$ and $s_b$ are different for the two growth types for the same cell in the lineage. As a result, $\Delta_d$ is also different. Hence their distributions are different for exponential and linear growth types. In contrast, the division-time for each cell is extended by $t/2$ to the left and $t/2$ to the right, irrespective of the growth type. Therefore, $\Gamma(\tau_d)$ is the same for linear and exponential growth types.

\subsection{Extra figures}

\begin{figure}[h!]
    \centering
    \includegraphics[width=1\linewidth]{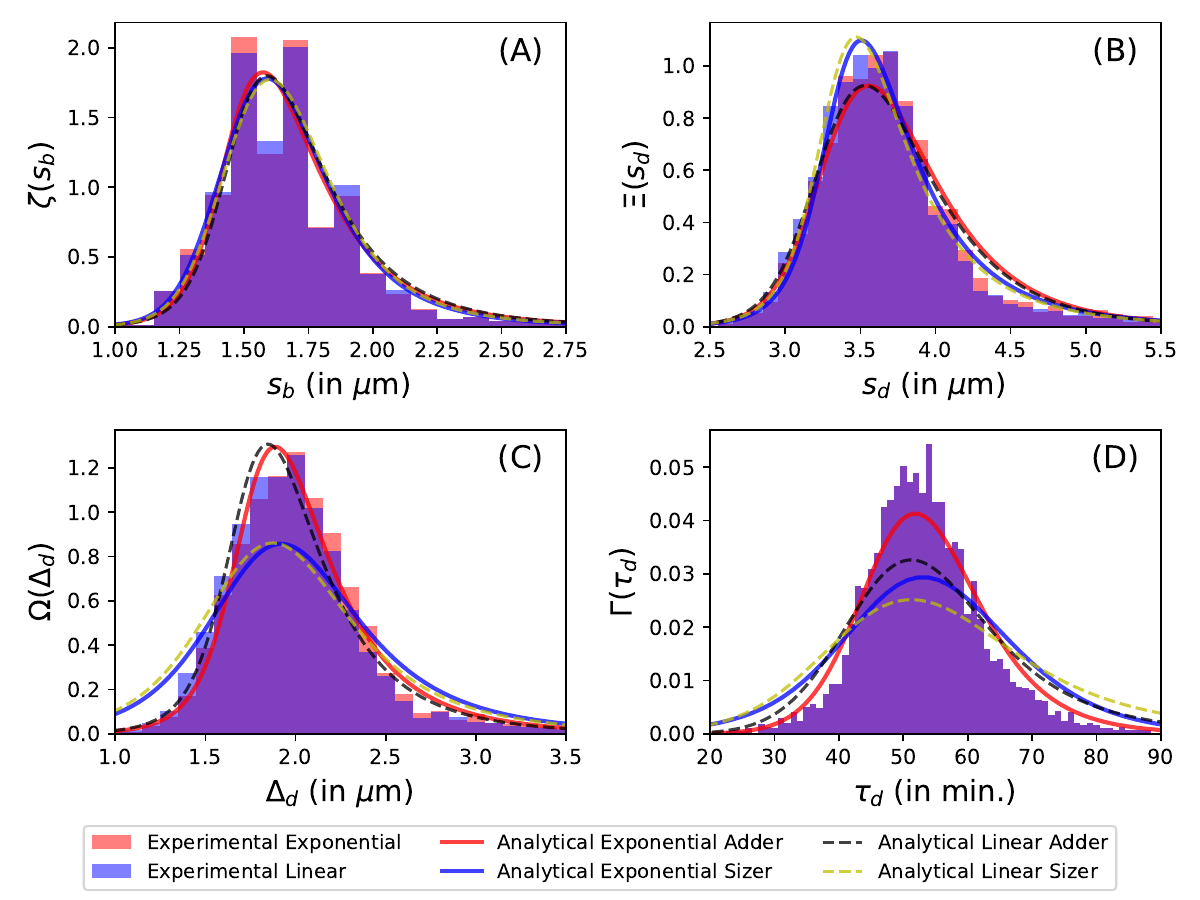}
\caption{Various experimental single-cell-level distributions (obtained from the single-cell experimental data for \textit{E. coli} ($27^{\circ}\text{C})$ \cite{tanouchi2017}) compared with their analytical results -- \textbf{(A)} Birth-size distribution, \textbf{(B)} Division-size distribution, \textbf{(C)} Division-added-size distribution, and \textbf{(D)} Division-time distribution, assuming a specific cell-division model (Sizer or Adder) and specific single-cell growth paradigm (linear or exponential).}
    \label{fig:tanouchiDisbns27C}
\end{figure}

\begin{figure}[h!]
    \centering
    \includegraphics[width=1\linewidth]{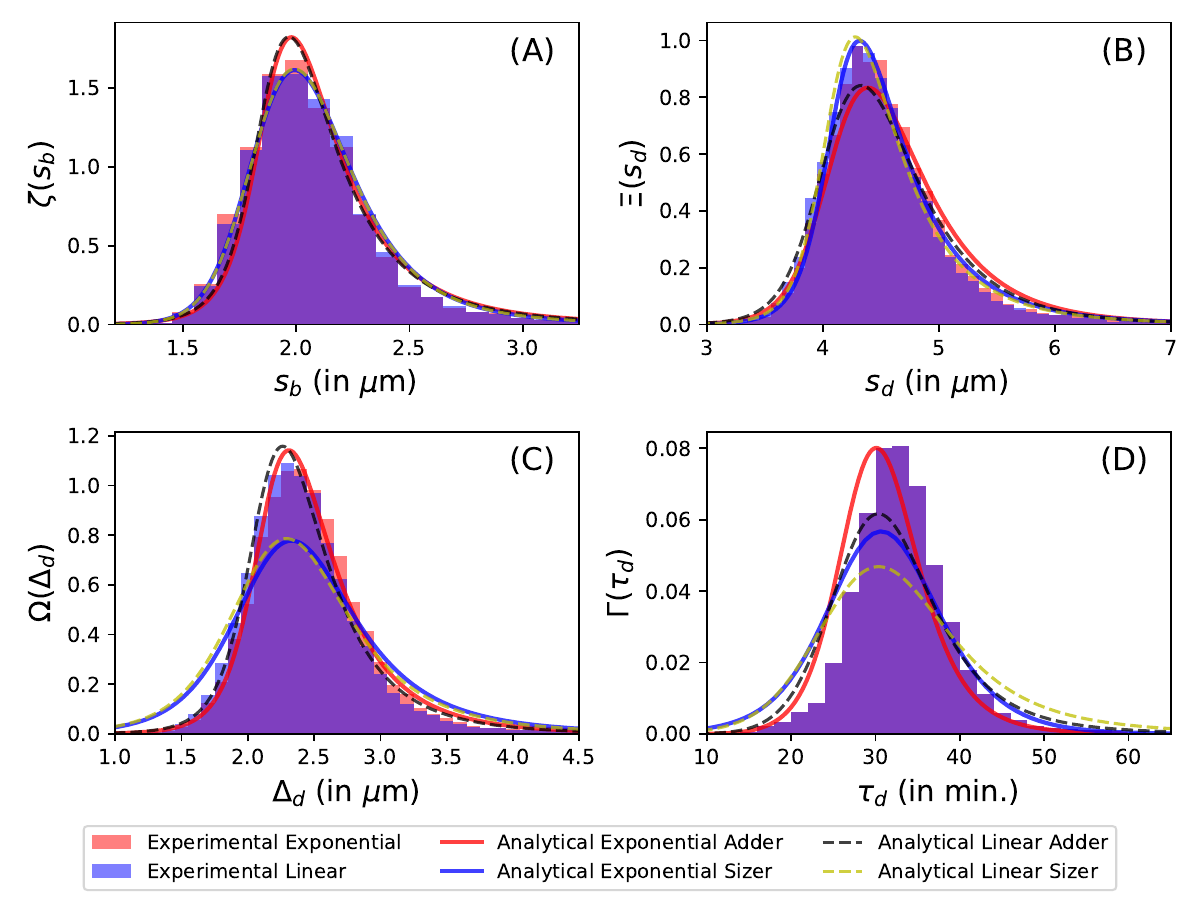}
\caption{Various experimental single-cell-level distributions (obtained from the single-cell experimental data for \textit{E. coli} ($37^{\circ}\text{C})$ \cite{tanouchi2017}) compared with their analytical results -- \textbf{(A)} Birth-size distribution, \textbf{(B)} Division-size distribution, \textbf{(C)} Division-added-size distribution, and \textbf{(D)} Division-time distribution, assuming a specific cell-division model (Sizer or Adder) and specific single-cell growth paradigm (linear or exponential).}
    \label{fig:tanouchiDisbns37C}
\end{figure}

\begin{figure}[h!]
    \centering
    \includegraphics[width=1\linewidth]{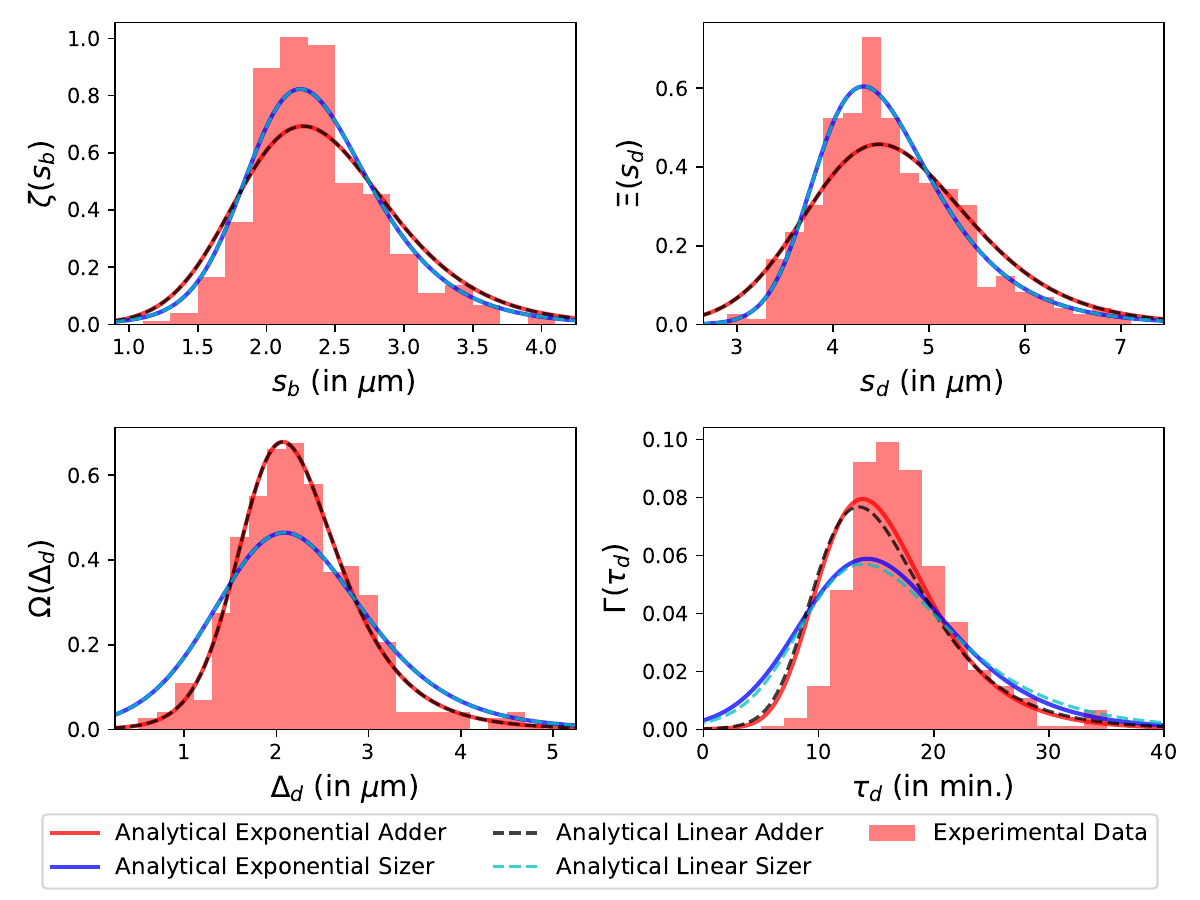}
\caption{Various experimental single-cell-level distributions (obtained from the single-cell experimental data for \textit{M. tuberculosis} labeled with SSB-GFP \cite{chungKarAmir2024}) compared with their analytical results -- \textbf{(A)} Birth-size distribution, \textbf{(B)} Division-size distribution, \textbf{(C)} Division-added-size distribution, and \textbf{(D)} Division-time distribution, assuming a specific cell-division model (Sizer or Adder) and specific single-cell growth paradigm (linear or exponential). }
    \label{fig:karDisbnsSSBGFP}
\end{figure}

\begin{figure}[h!]
    \centering
    \includegraphics[width=1\linewidth]{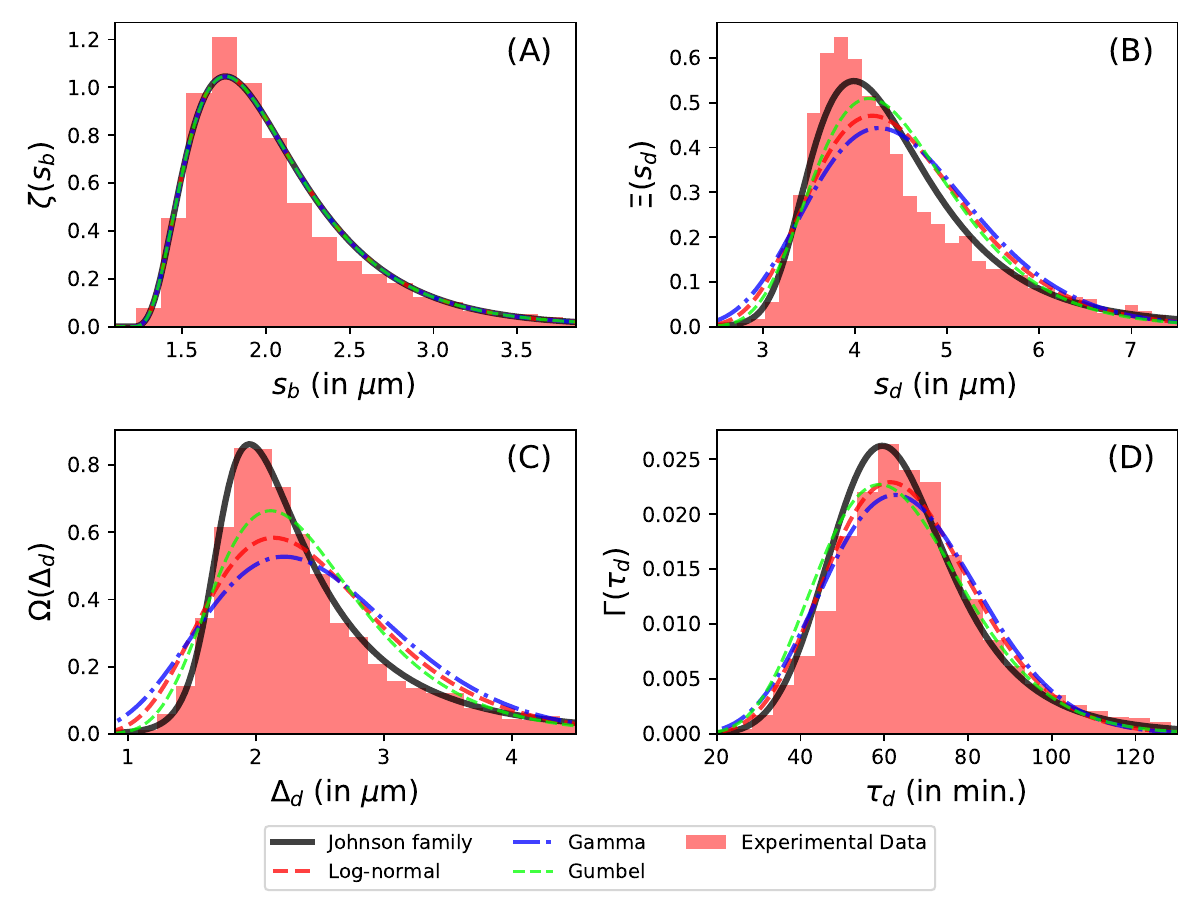}
\caption{Various single-cell-level distributions (for Adder model of cell division under single-cell exponential growth) -- \textbf{(A)} Birth-size distribution, \textbf{(B)} Division-size distribution, \textbf{(C)} Division-added-size distribution, and \textbf{(D)} Division-time distribution, compared with their analytical results corresponding to various cases when the principal distribution ($\Omega(\Delta_d)$) and growth-rate distribution ($\Lambda(\alpha)$) are taken to be the best fitting Johnson SU distribution, log-normal distribution, gamma distribution, or Gumbel distribution. $\zeta(s_b)$ for all of these cases is taken to be a Johnson SB distribution which has the values of its first four moments very close to the values calculated analytically using the recursive relation represented by Eq. \ref{eq: recursiveSbAdder}. The experimental corresponds to the data of \textit{E. coli} ($25^{\circ}\text{C})$ \cite{tanouchi2017}.}
    \label{fig:otherFits}
\end{figure}

\begin{figure}[h!]
    \centering
    \begin{tabular} {cc}
        \includegraphics[width=0.47\textwidth]{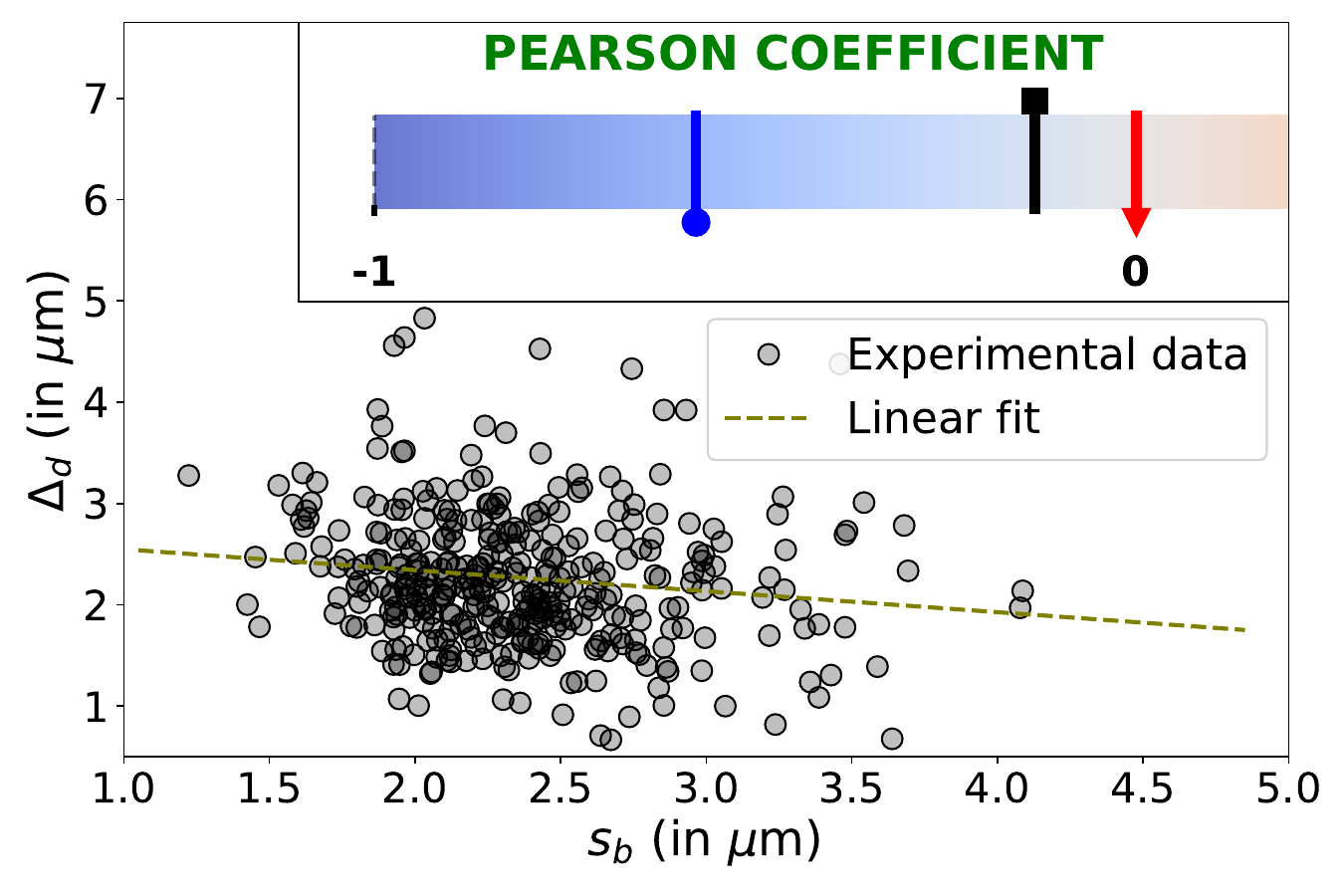}&
        \includegraphics[width=0.47\textwidth]{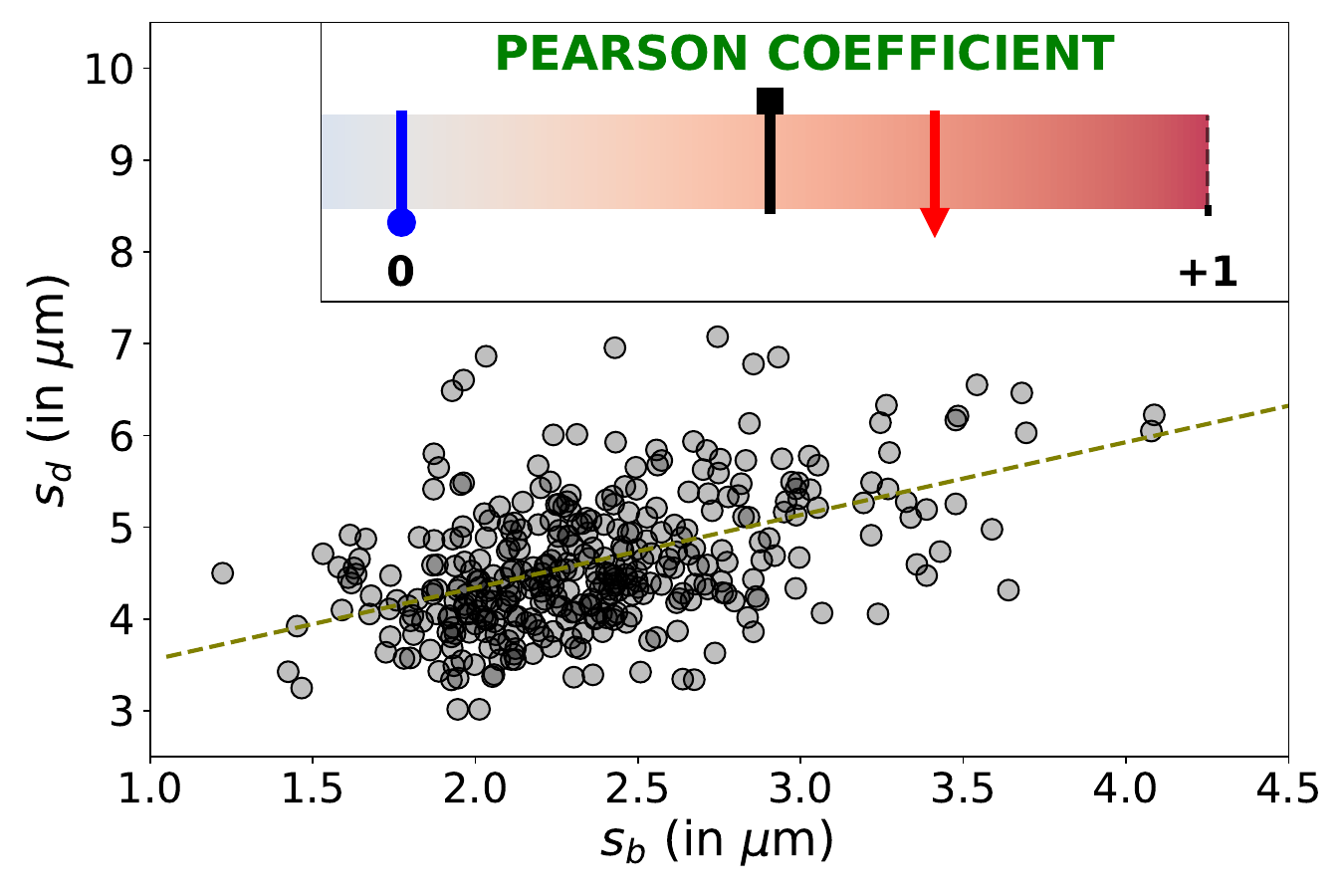}\\
    \end{tabular}
    \includegraphics[width=0.47\textwidth]{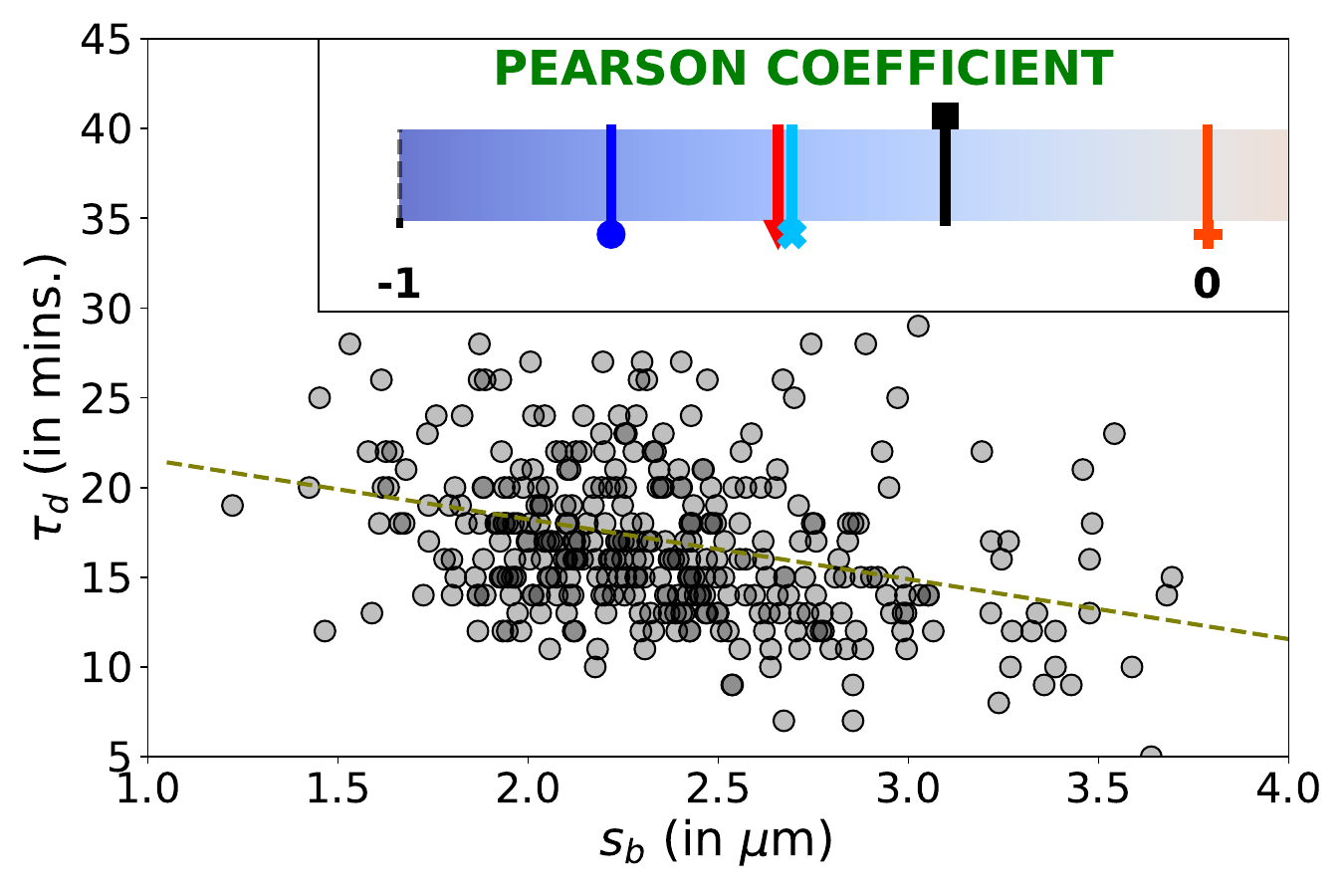}\\
    \caption{Various single-cell quantities ($\Delta_d$, $s_b$, and $\tau_d$) plotted against $s_b$ to infer their correlations with $s_b$. The data have been taken from Kar et al. \cite{chungKarAmir2024} (corresponding to SSB-GFP labeled CDC1551 strain of \textit{M. tuberculosis}). The insets in all of these plots compare Pearson Correlation Coefficients for the experimental data with its predicted analytical values for various growth and division models. The symbols used in the insets are the same for the insets in Fig. \ref{fig:correlations}.}
    \label{fig: correlationsKar}
\end{figure}

\begin{figure}[h!]
    \centering
    \includegraphics[width=0.82\linewidth]{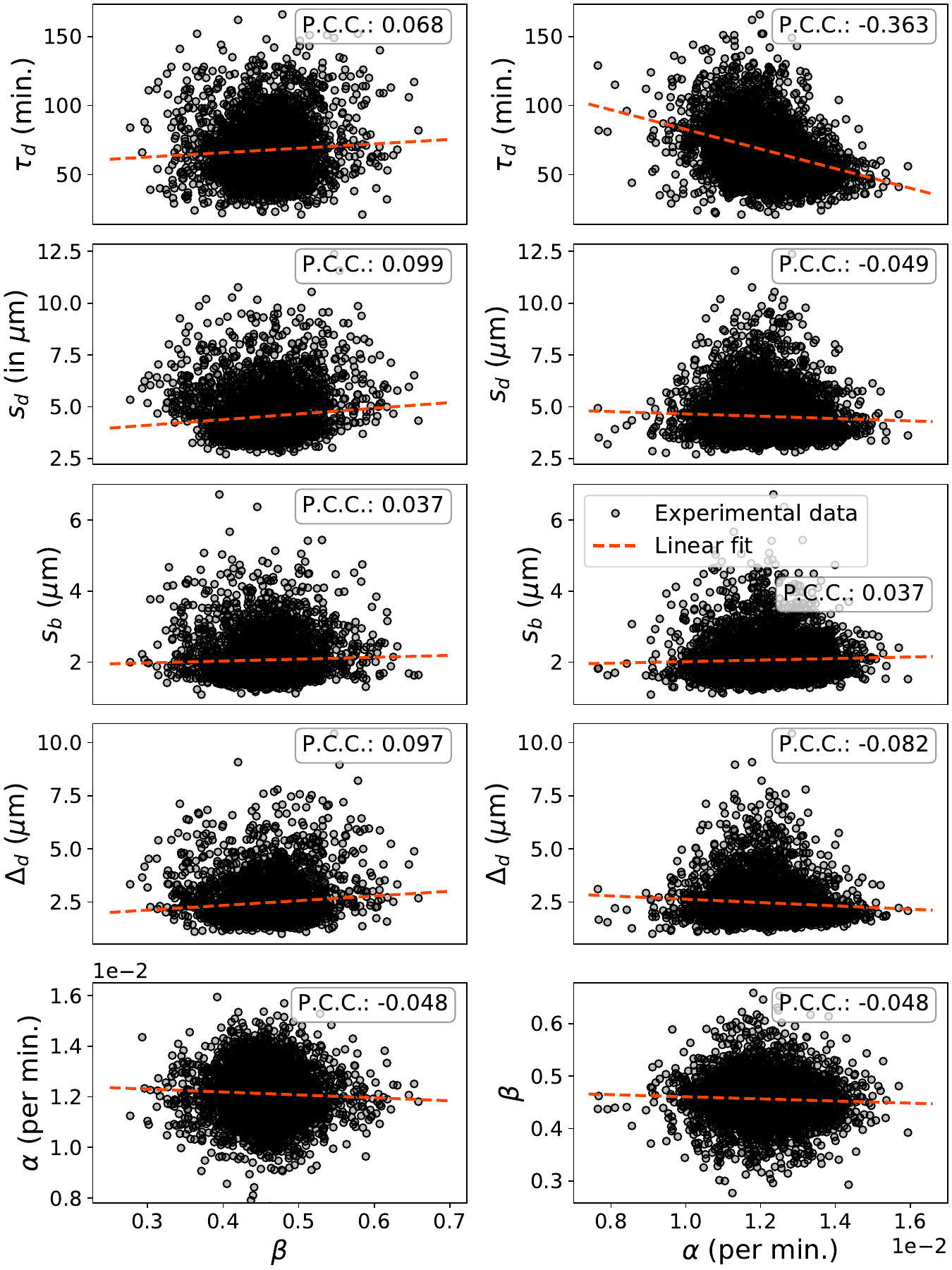}
\caption{Several single-cell quantities plotted against $\beta$ and $\alpha$ in order to show the correlations between them. The black dots correspond to the experimental data taken from \cite{tanouchi2017} (for \textit{E. coli}, $25^{\circ}\text{C}$). The dashed red line in the plots represent the linear fit to the data. ``P.C.C." inside the box in each sub-figure correspond to Pearson Correlation Coefficient. Note that the correlations of $\beta$ and $\alpha$ with other single-cell quantities are negligible, verifying our assumption regarding the independence between free parameters of our model, between size-related quantities and $\alpha$, and between size-related quantities and $\beta$. Interestingly, the correlation between $\tau_d$ and $\alpha$ is an exception, because it is not negligible. However, one can easily derive this correlation between $\tau_d$ and $\alpha$, and see that this correlation is equal to $-C(\alpha)/C(\tau_d)$ for both Sizer and Adder models, under both linear and exponential growth modes.}
    \label{fig:crossCorrelations}
\end{figure}

\begin{figure}[h!]
    \centering
    \begin{tabular} {cc}
        \includegraphics[width=0.47\textwidth]{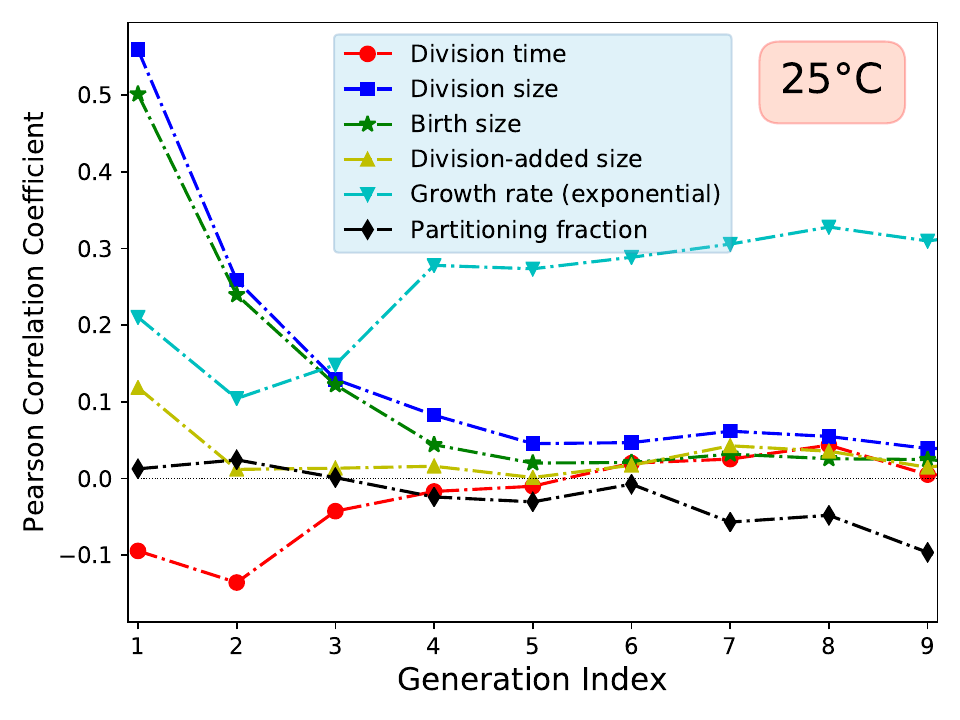}&
        \includegraphics[width=0.47\textwidth]{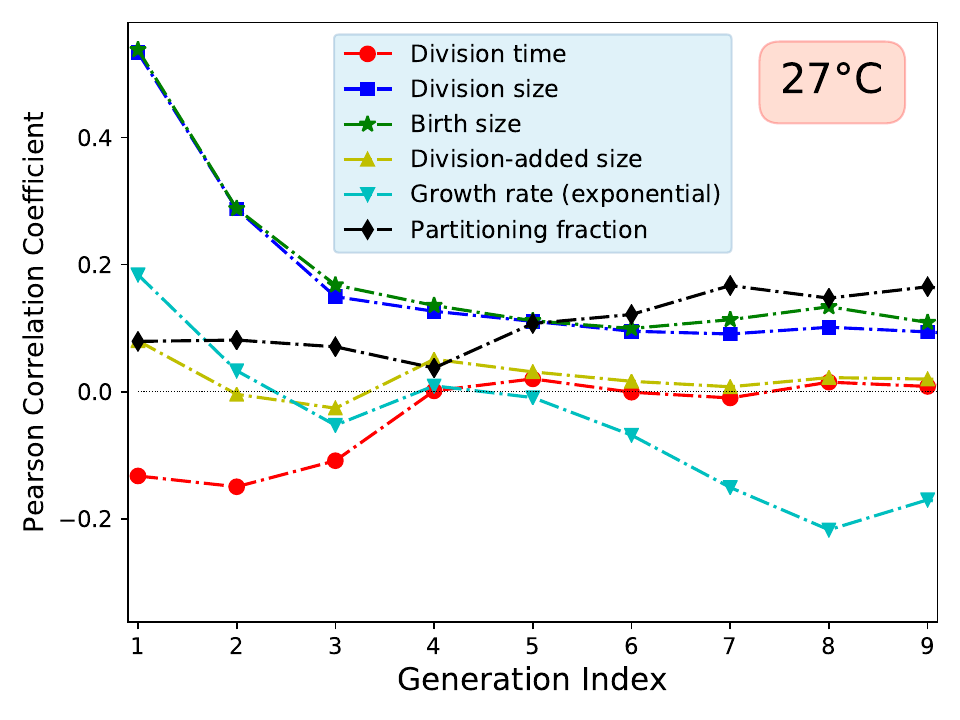}\\
    \end{tabular}
    \includegraphics[width=0.47\textwidth]{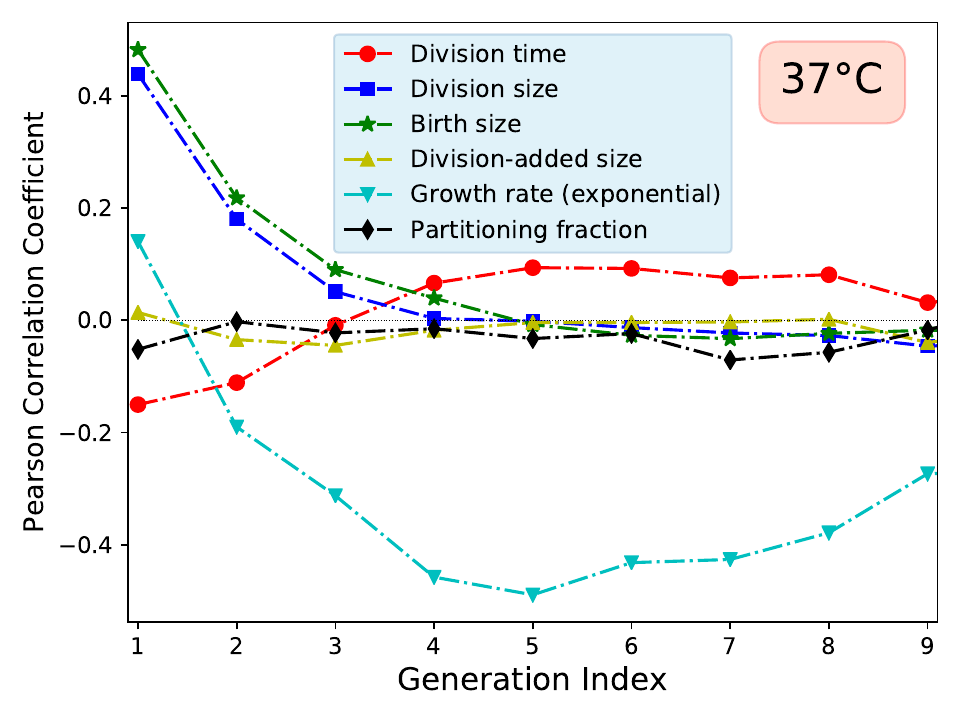}\\
    \caption{Correlations of single-cell quantities, such as $\tau_d$, $s_d$, $s_b$, $\Delta_d$, $\alpha$, and $\beta$, with their corresponding values in preceding generations within a lineage at the three different temperature reported in the study \cite{tanouchi2017} (for \textit{E. coli}, $25^{\circ}\text{C}$, $27^{\circ}\text{C}$, and $37^{\circ}\text{C}$). A symbol corresponding to Generation Index = $1$ represent the correlation between the single-cell quantity for the daughter cell and the corresponding quantity for the parent cell; a symbol corresponding to Generation Index = $2$ represent the correlation between the single-cell quantity for the daughter cell and the corresponding quantity for the grandparent cell. Similarly higher generation indices represent higher order autocorrelations for these quantities. The autocorrelations in $\beta$ and $\Delta_d$ are negligible. The autocorrelations for other quantities, such as $s_d$, $s_b$, and $\tau_d$, are noisy and only their overall decay can be discerned (these decaying patterns can be explained by the predictions of the Adder model as discussed by Jun et al. in the Supplemental Information \cite{taheriJun2015}).}
    \label{fig:autoCorrelations}
\end{figure}

\begin{figure}[h!]
    \centering
    \includegraphics[width=1\linewidth]{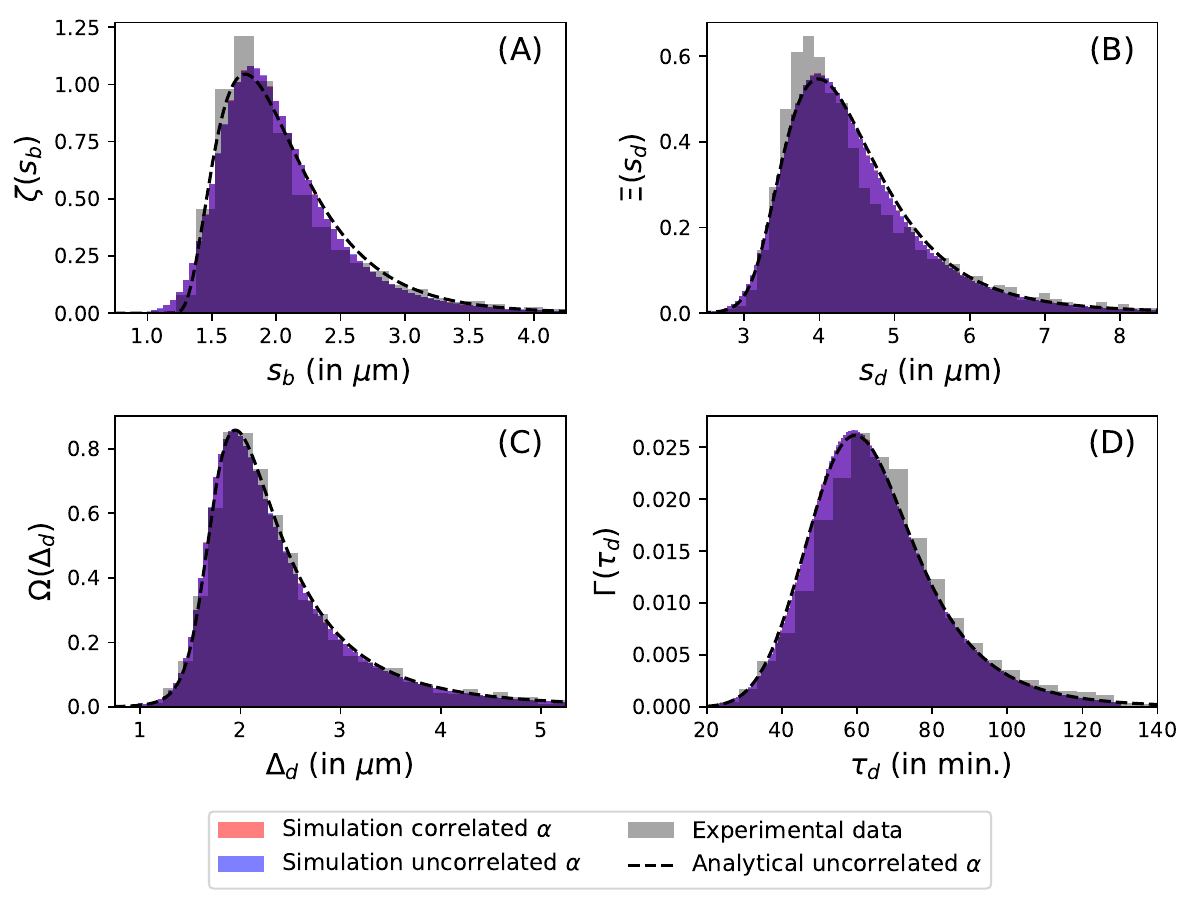}
\caption{Various single-cell-level distributions (for Adder model of cell division under single-cell exponential growth) -- \textbf{(A)} Birth-size distribution, \textbf{(B)} Division-size distribution, \textbf{(C)} Division-added-size distribution, and \textbf{(D)} Division-time distribution, compared for different cases of autocorrelations in growth rate with their corresponding experimental distributions. The experimental data corresponds to the single-cell data for \textit{E. coli} ($25^{\circ}\text{C})$ \cite{tanouchi2017}. ``Simulation correlated $\alpha$'' in the legend corresponds to the simulation results for the case when the auto-correlations in $\alpha$ are present. For this case, correlated samples of $\alpha$ were generated using a first-order autoregressive (AR(1)) process with the same probability distribution as for the case of uncorrelated $\alpha$. The lag-1 autocorrelation coefficient was set to (0.5), yielding an exponentially decaying autocorrelation function, $\rho(k)=0.5^k$. This makes the correlation (P.C.C.) between the growth rates of mother and daughter cells to be 0.5, the correlation between the growth rates of grandmother and daughter cells to be 0.25, and so on. The motivation to choose such decaying autocorrelations for $\alpha$ is the autocorrelations observed in Fig. \ref{fig:autoCorrelations} (although it is very noisy), and in previous study \cite{taheriJun2015}. ``Simulation uncorrelated $\alpha$'' corresponds to the simulation results for the case when the auto-correlations in $\alpha$ are absent. ``Analytical uncorrelated $\alpha$'' refers to the analytical distributions obtained for the case when auto-correlations in the growth rate are absent. The simulation results for the cases of correlated and uncorrelated $\alpha$ match. The slight mismatch between the analytical results and simulation results for $\zeta(s_b)$ reflects the fact that only first four moments were used in order to get the analytical form for this distribution, and therefore, the analytical result gives only the approximate behavior for $\zeta(s_b)$. However, the other single-cell-level distributions obtained analytically match with the simulation results, which shows that this approximation for $\zeta(s_b)$ (up to only first four moments) is valid in order to obtain the other distributions through probability transformations.}
    \label{fig:autoCorrelationsAlpha}
\end{figure}

\clearpage
\begin{acknowledgments}
Vikas gratefully acknowledges the University Grants Commission (UGC), India, for financial support through the Junior Research Fellowship Number 221610131832. A.R. acknowledges research support from IIT Delhi and ANRF-ECRG (Grant Number: ANRF/ECRG/2024/006991/LS). We would like to thank Monojit Chatterjee, Ambeesh Rolta, Dibyendu Das, and Sandeep Choubey for helpful discussions. Finally, we would like to thank the anonymous reviewers from previous submission for their critical comments which has now been incorporated in this manuscript.
\end{acknowledgments}

\section*{Author Contributions}
A.R. conceptualized the research; Vikas, R.M., and A.R. designed the research; Vikas performed the research; Vikas, R.M., and A.R. wrote the paper; R.M. and A.R. supervised the work.

\section*{Declaration of interests}
The authors declare no competing interests.

\bibliography{sources}

\end{document}